\documentclass[11pt]{elsarticle}
\usepackage{amsmath}
\usepackage{amsfonts}
\usepackage{amssymb}
\usepackage{amsthm}
\usepackage{bm}
\usepackage{indentfirst}
\usepackage{graphicx}
\usepackage{subfigure}
\usepackage{array}
\usepackage{fullpage}

\newcommand{\mo}{\mathcal{O}}
\DeclareMathAlphabet{\mathsfsl}{OT1}{cmss}{m}{sl}
\newcommand{\newtensor}[1]{\mathsfsl{#1}}
\renewcommand{\vec}[1]{\mathit{\boldsymbol{#1}}}
\newcommand{\dif}{\mathrm{d}}

\newcommand{\ba}{\bm\alpha}
\newcommand{\bx}{\bm\xi}
\newcommand{\ve}{\varepsilon}
\newcommand{\bmx}{\bar{\bm X}}
\newcommand{\PreserveBackslash}[1]{\let\temp=\\#1\let\\=\temp}
\newcolumntype{C}[1]{>{\PreserveBackslash\centering}p{#1}}
\newcolumntype{R}[1]{>{\PreserveBackslash\raggedleft}p{#1}}
\newcolumntype{L}[1]{>{\PreserveBackslash\raggedright}p{#1}}

\numberwithin{equation}{section}
\newtheorem{thm}{Theorem}[section]

\theoremstyle{definition}

\newtheorem{rem}[thm]{Remark}

\usepackage{titlesec}
\usepackage{graphicx}
\usepackage{subfigure}
\usepackage{array}
\usepackage{fullpage}
\usepackage{hyperref}
\usepackage{color,soul}
\usepackage{amsfonts,amssymb}
\usepackage{amsmath}
\usepackage{newfloat, algcompatible}
\usepackage{algcompatible}
\usepackage{etoolbox}

\DeclareFloatingEnvironment[fileext=loa, listname=List of Algorithms, name=ALGORITHM, placement=tbhp]{algorithm}

\begin{document}
\begin{frontmatter}

\title{Systematic parameter inference in stochastic mesoscopic modeling}

\author[pnnl]{Huan Lei}
\author[pnnl]{Xiu Yang}
\author[brown]{Zhen Li}
\author[brown]{George Em Karniadakis\corref{gk}}
\address[pnnl]{Pacific Northwest National Laboratory, Richland, WA 99352, USA}
\address[brown]{Division of Applied Mathematics, Brown University, Providence, RI 02912, USA}
\cortext[gk]{Corresponding author.}
\ead{george\_karniadakis@brown.edu}

%\maketitle
%%%%%%%%%%%%%%%%%%%%%%%%%%%%%%%%%%%%%
\begin{abstract}
We propose a method to efficiently determine the optimal coarse-grained 
force field in mesoscopic stochastic 
simulations of Newtonian fluid and polymer melt systems modeled by dissipative 
particle dynamics (DPD) and energy conserving dissipative particle
dynamics (eDPD). The response surfaces of various target properties (viscosity, diffusivity,
pressure, etc.) with respect to model parameters are constructed based on
the generalized polynomial chaos (gPC) expansion using simulation results on sampling
points (e.g., individual parameter sets). To alleviate the computational cost to 
evaluate the target properties, we 
employ the compressive sensing method to compute the coefficients
of the dominant gPC terms given the prior knowledge that the coefficients are
``sparse''. The proposed method shows comparable accuracy with the standard
probabilistic collocation method (PCM) while it imposes a much weaker
restriction on the number of the simulation samples especially for systems with
high dimensional parametric space. Fully access to the response surfaces
within the confidence range enables us to infer the optimal force parameters
given the desirable values of target properties at the macroscopic scale.
Moreover, it enables us to investigate the intrinsic relationship between the
model parameters, identify possible degeneracies in the parameter space, and
optimize the model by eliminating model redundancies. The proposed method
provides an efficient alternative approach for constructing mesoscopic models
by inferring model parameters to recover target properties
of the physics systems
(e.g., from experimental measurements), where those force field parameters
and formulation cannot be derived
from the microscopic level in a straight forward way.
\end{abstract}

\begin{keyword}
Coarse-grained force field, dissipative particle dynamics,
energy conserving dissipative particle dynamics, compressive sensing,
generalized polynomial chaos, model reduction, high dimensionality
\end{keyword}
\end{frontmatter}
%%%%%%%%%%%%%%%%%%%%%%%%%%%%%%%%%%%%%%%%
\section{Introduction}
\label{sec:intro}
It is well known that many of the macroscopic properties observed in soft matter
systems, such as liquid crystals, polymers, and colloids are natural
consequences of the physical processes at the microscopic level. Accurate
modeling of the corresponding processes enables us to successfully predict
the properties of these material systems as well as, in turn, calibrate the
parameters of the models.
%Therefore, it is of great importance in both
%scientific and industrial aspects \cite{Material_Genome}.
Molecular dynamics (MD) simulation, in conjunction with optimized force fields,
has been  successfully applied to study various
physical systems \cite{Jorgensen_JACS_1996,Spyriouni_JACS_1999}. However, due to
the explicit modeling of individual atomistic particles, the MD simulation
method is limited and cannot reach the spatial and temporal scales relevant to the
collective motion we are interested in, i.e., at the mesoscale.

To overcome this limitation, an alternative approach is to coarse grain (CG) the
system by representing several atomistic particles as a single virtual particle.
The essential idea is to eliminate the fast modes and corresponding degrees of freedom (DOF)
while only keeping those DOFs relevant to the scale of our interest, represented
as the CG particles. The static properties of the CG system are closely related
to the governing force field, which have been studied extensively
\cite{EspSer97,KlaDie04,LouBol00,KinHyo07Mol,AkkBri01,HarAdh06,FukTak02}.
Espan\~{o}l~\cite{EspSer97} modeled the DPD particles by grouping
several Lennard Jones (LJ) particles into clusters, and derived the conservative
force field from the radial distribution function of the clusters. Bolhuis
\textit{et al.}~\cite{LouBol00} mapped the semi-dilute polymer solution onto soft
particles via an effective pairwise potential. Kremer \textit{et al.}
~\cite{HarAdh06} and Fukunaga \textit{et al.}~\cite{FukTak02} extracted the
effective force field for complex polymers from the distribution functions of
the bond length, bending angle and torsion angle. By carefully choosing the CG
sites (``super atoms'') and tuning the CG force field parameters, these CG
models can reproduce the specified structural properties of the atomistic
systems.

However, currently there are still two open
questions for constructing the CG model and force field. First, most of
the CG force fields are constructed by targeting certain static properties
(e.g., the pair and angle distribution functions) of the atomistic system,
while the constructed CG force fields show limitation if we consider other
static properties (e.g., the equation of state) \cite{AkkBri00, Lei_Cas_2010}.
In particular, how to construct a CG model with a thermodynamically
consistent force field is still an open question. Moreover, the CG force fields
are insufficient to reproduce the dynamic properties of the atomistic system.
Instead, two additional (dissipative and random) force terms should be
introduced to compensate for the eliminated atomistic DOFs during the CG
procedure \cite{KinHyo07}. This is represented by the force terms in
mesoscopic simulation methods such as dissipative particle dynamics (DPD)
\cite{Hoogerbrugge_SMH_1992,Espanol_SMO_1995}
{ and its variations, e.g., energy-conserving dissipative particle
dynamics (eDPD) \cite{QiaoH07,LiTLBK14}}. Computing the dissipative and
random force terms directly from the atomistic systems is a non-trivial task.
Eriksson \textit{et al}. \cite{EriNil08} estimated the dissipative
force term by the force covariance function, whereas Lei \textit{et al.}
\cite{Lei_Cas_2010}, Hij{\'{o}}n \textit{et al}. \cite{Espanol_Farady_2010}, 
Izvekov \textit{et al}. \cite{Izvekov_Rice_JCP_2014} and 
Li \textit{et al}. \cite{Li_Kar_SM_2014} computed
the dissipative force term using the Mori-Zwanzig formulation
\cite{Mor65,Zwa60}. While the computed dissipative force terms successfully
reproduce the dynamic properties in the dilute and semi-dilute regime, the force
terms show limitation in reproducing the dynamic properties for highly correlated
systems. Currently, it is still unclear how to determine the optimized CG force
terms with correct dynamic properties at high density regime.

To circumvent those difficulties discussed above, we study the CG systems by
considering an alternative question: if we
are only interested in \textit{some} particular properties (e.g., \textit{target} properties)
related to the physical system, how do we choose the optimal CG modeling parameters within certain
confidence range? That is, how to calibrate the CG modeling parameters
for a specific set of target properties observed at the
macroscopic scale?
We emphasize that the inferred parameters may not correspond to the values in
mesoscopic models directly constructed from the microscopic systems in \textit{ab initio}
way. However, these parameters should be sufficient to recover the target
properties we are interested in if the CG model is constructed appropriately.
In this sense, this study also enables us
to validate the proposed CG model as well as to analyze the sensitivity of
the macroscopic quantities on individual modeling parameters.
%\textcolor{green}{For example, Rizzi \textit{et al.}
%\cite{Najm_UQ_2012_a, Najm_UQ_2012_b}
%%used the generalized polynomial chaos (gPC)
%%\cite{GhanemS91, XiuK02} expansion to construct a surrogate model and
%inferred the three-dimensional force parameters in TIP4P water model by using
%the macroscopic
%properties. Koumoutsakos \textit{et al.} \cite{Pan_Kou_JCP_2012}
%%implemented the transitional Markov chain Monte Carlo (TMCMC) algorithm to
%investigated the MD system of liquid and gaseous argon.}
%%by adopting the kriging algorithm as the adaptive surrogate model.}
%(\textcolor{green}{remark by Huan: I have rewritten
%this paragraph by emphasizing parameter inference rather than surrogate model.
%I think it is better
%to remove these examples to the text where we discuss about constructing surrogate
%model}\textcolor{blue}{remark by Xiu: I removed sentences with ``surrogate
%model" and put the surrogate model in the next paragraph} )

In this work, we aim to investigate the static and dynamic properties of
mesoscopic systems governed by the DPD/eDPD force field within a high dimensional
random parameter space. In particular, we study the
dynamic properties of two mesoscopic systems:  a
non-Newtonian polymer melt system governed by DPD force field 
within a 6-dimensional parameter space; and a non-isothermal model for
liquid water at different temperatures governed by eDPD force field within
a 4-dimensional parameter space.
In order to calibrate the parameters in mesoscopic systems, we employ
the Bayesian framework:
\[\pi(\bm\theta)\propto \mathcal{L}(\bm\theta, \bm P^t)q(\bm\theta),\]
where $\bm\theta$ are parameters to infer, $\bm P^t$ are target
properties, $\pi$ is the posterior distribution, $\mathcal{L}$ is the
likelihood function, and $q$ is the prior distribution. The selection of parameters
$\bm\theta$ relies on the posterior distribution $\pi$, which is usually
obtained by the Markov Chain Monte Carlo (MCMC) method \cite{gelman2013bayesian, liu2008monte} or its variants. However,
the MCMC method usually requires evaluating the likelihood function thousands of
times, which in turn requires running the costly DPD simulation thousands of times.
This requirement is prohibitive for complex systems, hence an accurate and fast
\textit{surrogate model} (which approximates the response surface) is necessary.
In this work, we employ generalized polynomial chaos (gPC) \cite{GhanemS91, XiuK02}
to build the surrogate model for DPD systems in the form of a linear combination
of a set of special basis functions defined in the parameter space. This idea
has been used in solving inverse problems (e.g.,
\cite{MarzoukNR07, MarzoukN09}) and MD modeling of water systems (e.g.,
\cite{Najm_UQ_2012_a, Najm_UQ_2012_b}). This gPC surrogate model facilitates
the implementation of MCMC and dramatically accelerates the evaluation of the
likelihood function $\mathcal{L}$, hence enabling us to infer the parameters
efficiently.

The construction of surrogate models can be achieved by functional representations employed in
uncertainty quantification (UQ), e.g., (adaptive) sparse grid method
\cite{XiuH05, GanapathyZ07, FooWK08, NobileTW08, MaZ09} or (adaptive) ANOVA
method \cite{MaZ10, FooK10, ZhangCK12, YangCLK12}. All these methods are
categorized as probabilistic collocation methods (PCM) as they provide smart
strategies to select sampling points and weights.
%However, in practice, the
%curse of high-dimensionality is still a big issue. Also, the number of
%simulations is constrained due to the complexity of the system and limited
%computational source.
In recent years, efforts have been made to propose a non-adaptive but
simple and accurate method for high dimensional problems
\cite{Li10, DoostanO11, YanGX12, YangK13, Lei_Yang_MMS_2015, Yang_Lei_JCP_2016} when the system is \textit{sparse},
which within the UQ framework means that only a small portion of the gPC coefficients
has relatively large values while the others are close to zero. This idea stems from the
compressive sensing field \cite{CandesT05, CandesRT06, DonohoET06, BrucksteinDE09}. More
precisely, if we know \textit{a priori} that the vector of gPC coefficients is sparse, we
can employ compressive sensing techniques to ``recover" these coefficients
accurately using only a few (deterministic) simulation data. This method can be
considered as post processing of the sampling method, e.g., Monte Carlo method or PCM method, hence it is
flexible as we can always incorporate new available data and obtain better
estimates. This method has an advantage over many popular methods, e.g.,
the standard sparse grid method, which is popular in UQ studies today, as 
it requires a fixed number of additional simulations to reach the next accuracy level.
This requirement is prohibitive in
practice when the dimension of the problem is high. Even with the adaptive
method, the selection of adaptivity criteria can be constrained by the
limitation of computational resources. Moreover, in MD and DPD simulations, due
to the existence of the thermal noise, it is difficult to obtain a highly
accurate gPC surrogate model. In other words, even if we can afford running
simulations with a high order PCM method, we may not be able to reduce the error
of the surrogate model. This is different from research on UQ in solving
stochastic PDEs, where high accuracy can be expected with high
order methods. Therefore, it is more
helpful to adopt a method which can efficiently construct
the gPC expansion with small errors (ideally close to the thermal noise level).

\begin{figure}[!h]
\centering
\setlength{\unitlength}{1.2mm}
\begin{picture}(130, 130)
\linethickness{1pt}
\put(10,106){\framebox(48,18){\shortstack{Design material with desirable \\
  macroscopic properties $\bm P^t$, e.g., \\
  viscosity, shear modulus, etc. via \\
  DPD model with parameter set $\bm\theta$}}}
\put(34,106){\vector(0,-1){30}}
\put(10,64){\framebox(48,12){\shortstack{Run DPD simulations
 to obtain \\ macroscopic properties $\bm P^{DPD}$}}}
\put(34,64){\vector(0,-1){22}}
\put(34,42){\line(-2,-1){24}} \put(34,42){\line(2,-1){24}}
\put(34,18){\line(-2,1){24}} \put(34,18){\line(2,1){24}}
\put(18,26){\shortstack{Compare $\bm P^{DPD}$ and $\bm P^{t}$:\\
            $\Vert \bm P^{DPD} - \bm P^t\Vert<\epsilon$?}}
\put(58,30){\line(1,0){46}} \put(104,30){\vector(0,1){12}}
\put(80,42){\framebox(48,25){\shortstack{Run DPD simulations with\\
different sets of $\bm \theta$ to obtain \\ samples $\bm P^s$ and
construct \\ \textit{gPC} expansion as surrogate \\ model based on $\bm P^s$ via\\
    \textit{compressive sensing}}}}
\put(104,67){\vector(0,1){15}}
\put(80,82){\framebox(48,18){\shortstack{Implement Bayesian inference\\  based
 on $\bm P^t$ and the surrogate \\ model to obtain PDF of $\bm\theta$ then \\
     select optimal $\bm\theta$}}}
\put(80,91){\vector(-1,0){46}}
\put(36,12){Yes} \put(76,31){No}
\put(34,18){\vector(0,-1){10}}
\put(10,0){\framebox(48,8){\shortstack{Finish}}}
\end{picture}
\caption{Overview of estimating parameters to achieve desired material
properties using the DPD model.}
\label{fig:dpd}
\end{figure}
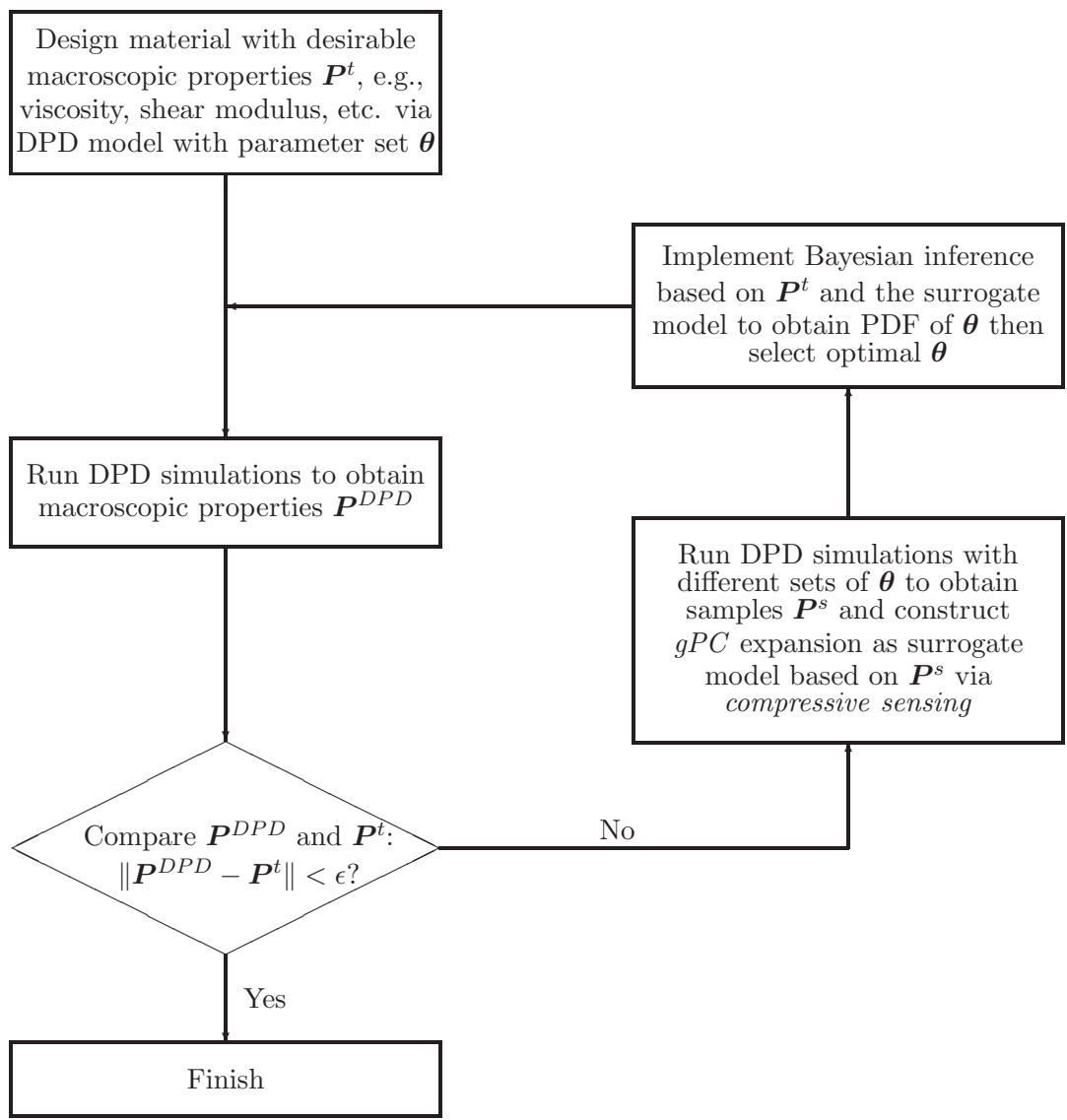

After obtaining a good surrogate model of the response surface
of the target property, we employ the Bayesian inference to explore suitable
model parameters in DPD.
%With the gPC expansion, we are able to dramatically
%reduce the cost of evaluating the likelihood, since we do not need to run the
%whole DPD simulation when this evaluation is needed. This type of gPC
%based Bayesian inference has been applied in several inverse problems, e.g.,
%\cite{MarzoukNR07, MarzoukN09}, and we adopt a similar framework but employ
%a new technique to construct the gPC expansion. Thus,
For material
design problems, given the desirable target properties $\bm P^t$, we can
firstly build a DPD model with empirical (commonly used) constants. Then, we parametrize these
constants to obtain a parameter set $\bm\theta$. Next, we run the DPD simulation
with different sets of $\bm\theta$ values to obtain samples of target properties
$\bm P^S$ and build the gPC approximation (surrogate model) of target properties
depending on $\bm\theta$ via compressive sensing. Finally, we implement the Bayesian
inference based on $\bm P^t$ and the gPC expansion to infer
the suitable $\bm\theta$
for the DPD model. The above procedure can be repeated until the DPD model
reaches the required accuracy by increasing the accuracy of the surrogate model
or by modifying the range of the parameters. Figure \ref{fig:dpd} provides an
overview of the entire procedure of inferring the parameters in the DPD model
via gPC expansion and compressive sensing. The difference between $\bm P^t$ and
$\bm P^{DPD}$ can be measured by the summation of relative error of each target
property.

%-------------------------------------------------------------------------------

This paper is organized as follows. Section \ref{sec:dpd} is devoted to a
brief introduction of the mesoscopic simulation method and models
considered in the present study, i.e., the non-Newtonian polymer melt model
based on dissipative particle dynamics (DPD) and the non-isothermal liquid
water model based on the /energy-conserving dissipative particle
dynamics (eDPD). In Section \ref{sec:gpc_l1},
we introduce the compressive sensing method and its application in the gPC
coefficient computation. In Section \ref{sec:result}, we construct the gPC
expansions of various target properties over the parameter space for both
the polymer melt and the liquid
water systems. Using the gPC expansion as the surrogate model, we estimate the
optimal force parameters and eliminate possible model redundancies with the
prescribed value of the target properties via Bayesian inference. In
Section \ref{sec:discussion}, we summarize our work and briefly discuss
future directions.

%===============================================================================
%===============================================================================
%===============================================================================

\section{Simulation method and physical model}
\label{sec:dpd}
%-------------------------------------------------------------------------------

\subsection{Dissipative Particle Dynamics}
Dissipative Particle Dynamics (DPD) \cite{Hoogerbrugge_SMH_1992,Groot_DPD_1997}
is a particle based mesoscopic simulation method. Each DPD particle represents a
coarse-grained virtual cluster of multiple atomistic particles
\cite{Lei_Cas_2010}.
In standard DPD formulation \cite{Hoogerbrugge_SMH_1992}, the motion of each
particle is governed by
%-------------------------------------------------------------------------------
\begin{equation}
\begin{split}
\dif\bm r_i &= \bm v_i\dif t \\
\dif\bm v_i &= (\bm F^C_i\dif t + \bm F^D_i\dif t + \bm F^R_i\sqrt{\dif t})/m,
\end{split}
\label{eq:DPD_eq1}
\end{equation}
%-------------------------------------------------------------------------------
where $\bm r_i$, $\bm v_i$, $m$ are the position, velocity, and mass of the
particle $i$, and $\bm F^C_i$, $\bm F^D_i$, $\bm F^R_i$ are the total
conservative, dissipative and random forces acting on the particle $i$,
respectively. Under the assumption of pairwise interactions the DPD forces can
be decomposed into pair interactions with the surrounding particles by
%-------------------------------------------------------------------------------
\begin{equation}
\label{eq:DPD_eq2}
\begin{aligned}
& \bm F_{ij}^C =
\begin{cases}
a(1.0 - r_{ij}/r_c) \bm e_{ij}, & r_{ij} < r_c,\\
    0, &r_{ij} > r_c,
\end{cases}\\
& \bm F_{ij}^D = -\gamma w_{D}(r_{ij})(\bm v_{ij}\cdot\bm e_{ij})\bm e_{ij},\\
& \bm F_{ij}^R = \sigma w_{R}(r_{ij})\zeta_{ij}\bm e_{ij},
\end{aligned}
\end{equation}
%-------------------------------------------------------------------------------
where $\bm r_{ij}=\bm r_i-\bm r_j$, $r_{ij} =|\bm r_{ij}|$, $\bm e_{ij} =
\bm r_{ij}/r_{ij}$, and $\bm v_{ij}=\bm v_i-\bm v_j$. $r_c$ is the cut-off
radius beyond which all interactions vanish. The coefficients $a$, $\gamma$ and
$\sigma$ represent the strength of the conservative, dissipative and random
force, respectively. The dissipative and random force terms are coupled with the
system temperature by the fluctuation-dissipation theorem
\cite{Espanol_SMO_1995} as $\sigma^2 = 2\gamma k_BT$. Here,
$\zeta_{ij}$ are independent identically distributed (\textit{i.i.d.}) Gaussian
random variables with zero mean and unit variance. The weight functions
$w_{D}(r)$ and $w_{R}(r)$ are defined by
%-------------------------------------------------------------------------------
%\begin{equation}
%\label{eq:DPD_eq3}
%\begin{split}
%w^{D}(r_{ij}) &= \left[w^R(r_{ij})\right]^2, \\
%w^{R}(r_{ij}) &= (1.0 - r_{ij} /r_c)^{k},
%\end{split}
%\end{equation}

\begin{equation}
\label{eq:DPD_eq3}
\begin{split}
w_{D}(r_{ij}) &= \left[w_R(r_{ij})\right]^2 \\
w_{R}(r_{ij}) &=
\begin{cases}
(1 - r_{ij} /r_c)^{k}, & r_{ij} < r_c\\
0, & r_{ij} > r_c\\
\end{cases}
\end{split}
\end{equation}
%-------------------------------------------------------------------------------
where $k$ is a parameter that determines the extent of dissipative and random
force envelopes.

While this method was initially proposed \cite{Hoogerbrugge_SMH_1992} to
simulate the complex hydrodynamic processes of isothermal fluid systems, the
particle based feature enables us to easily incorporate additional physical
models and extend its application to various soft matter and complex fluid
systems, such as polymer and DNA suspensions \cite{Spen00,FanPhan06,VsyCas05},
platelet aggregation \cite{PivRic09}, microgel \cite{Brown_Sta_NM_2014}, colloid 
suspension \cite{BoeCov97}, droplet wetting \cite{Li_Hu_POF_2013} and blood flow systems
\cite{PivKar08, Fedosov_RBC_2010,Fedosov_PNAS_2010, Lei_Kar_BJ_2012, Lei_Kar_PNAS_2013}. However, we note that the
intrinsic relationship between DPD and the atomistic force field is not well
understood yet. The force parameters of the mesoscopic models are usually chosen
empirically (referred as a specific ``point'' within parameter space) while the
variation of the macroscopic properties over the global parameter space has not
been fully studied. Alternatively, in this work, we simulate the 
non-Newtonian polymer melt systems by choosing the force parameters
varying within a range of empirical values; moreover, we aim to systematically
investigate the \textit{sensitivity} of the dynamic viscosity on individual
force parameters.

\subsection{Energy-conserving Dissipative Particle Dynamics}
In addition to the conservation of momentum in the DPD model
(Eq.~\eqref{eq:DPD_eq2}), the eDPD model also takes into account the
conservation of energy, which is described by the following equation
\cite{Espanol97eDPD,QiaoH07}:
\begin{equation}\label{eq:eDPD1}
  C_v\frac{\mathrm{d} T_i}{\mathrm{d} t}= q_{i} = \sum_{i\neq j}(q_{ij}^{C}+q_{ij}^{V}+q_{ij}^{R}) \ ,
\end{equation}
where $T_i$ is the temperature, $C_v$ is the thermal capacity of eDPD particles and
$q_i$ is the heat flux between particles. In particular, the three components of $q_i$ including the collisional heat flux $q^C$, viscous heat flux $q^V$, and random heat flux $q^R$ are given by
\cite{QiaoH07,LiTLBK14}:
%-------------------------------------------------------------------------------
\begin{align}
  & q_i^C = \sum_{j \ne i} k_{ij} \omega_{CT}(r_{ij}) \left( \frac{1}{T_i} - \frac{1}{T_j} \right) \ , \label{equ:heat1}\\
  & q_i^V = \frac{1}{2 C_v}\sum_{j \ne i}{ \left\{ \omega_D(r_{ij})\left[ \gamma_{ij} \left( \mathbf{e}_{ij} \cdot \mathbf{v}_{ij} \right)^2 - \frac{\left( \sigma _{ij} \right)^2}{m}\right] - \sigma _{ij} \omega_R(r_{ij})\left( \mathbf{e}_{ij} \cdot \mathbf{v}_{ij} \right){\zeta _{ij}} \right\} } \ , \label{equ:heat2}\\
  & q_i^R = \sum_{j \ne i} \beta _{ij} \omega_{RT}(r_{ij}) d {t^{ - 1/2}} \zeta _{ij}^e  \ , \label{equ:heat3}
\end{align}
%-------------------------------------------------------------------------------
where $k_{ij}$ and $\beta_{ij}$ determine the strength of the collisional and
random heat fluxes. The parameter $k_{ij}$ plays a role of thermal
conductivity and is given by $k_{ij}=C_v^2\kappa(T_i+T_j)^2/4k_B$ in which
$\kappa$ is interpreted as mesoscale heat friction coefficient
\cite{Marisol98,QiaoH07,He08,Eiyad10,LiTLBK14}, and $\beta_{ij}^2=2k_Bk_{ij}$. The weight
functions $\omega_{CT}(r)$ and $\omega_{RT}(r)$ in Eqs. $($\ref{equ:heat1}$)$
and $($\ref{equ:heat3}$)$ are given by
$\omega_{CT}(r)=\omega_{RT}^2(r)=\left(1-r/r_C\right)^{s}$ where $s$ is
the exponent of the weight functions. In our previous work \cite{LiTLBK14}, we
analyzed the sensitivity of transport properties to different eDPD parameters and found
that making $s$ a function of temperature is the best option for 
modeling the temperature-dependent properties of simple fluids
such as water and ethanol. However, obtaining an optimal functional form of $s$
so that an eDPD model can generate correct transport properties at various temperatures over
a wide range is a non-trivial task. In principle, however, this is one 
of the parameters that can be computed using our framework based on some target 
macroscopic properties, which will be 
demonstrated in sections \ref{subsec:edpd_model} and \ref{sec:eDPD}.

\subsection{Polymer model and parameter uncertainty}
\label{subsec:poly_uq}
The polymer melt system is modeled by $N_p$ flexible polymer chains in a domain
of $50 \times 20 \times 10$ with periodic boundary conditions with total DPD
particle number density $n = 3.0$. Each polymer chain consists of $N_b = 2\sim 5$
DPD particles connected by the Finitely Extensible Non-Linear Elastic (FENE)
potential given by
%-------------------------------------------------------------------------------
\begin{equation}
U_{FENE}=-\frac{k_s}{2}r^2_{max}
\log\left[ 1-\frac{|{\bm r}_i - {\bm r}_j|^2}{r^2_{max}}\right],
\label{eq:fene}
\end{equation}
%-------------------------------------------------------------------------------
where $k_s$ is the spring constant. The spring extension $r$ is limited by its
maximum value $r_{max}$ attained when the corresponding spring force becomes
infinite. The force parameters of the polymer melt system are given by
%-------------------------------------------------------------------------------
\begin{equation}
 \begin{aligned}
  a(\xi_1) &= \left<a\right> + \sigma_{a}\xi_1, & \gamma(\xi_2) &= \left<\gamma\right> + \sigma_{\gamma}\xi_2,\\
  k(\xi_3) &= \left<k\right> + \sigma_{k}\xi_3, & r_c(\xi_4) &= \left<r_c\right> + \sigma_{r_c}\xi_4,\\
  k_s(\xi_5)& = \left<k_s\right> + \sigma_{k_s}\xi_5, & r_{max}(\xi_6) &= \left<r_{max}\right> + \sigma_{r_{max}}\xi_6
 \end{aligned}
 \label{eq:para_uncertain}
 \end{equation}
%-------------------------------------------------------------------------------
where $\xi_1,...,\xi_6$ are \textit{i.i.d.} random variables uniformly distributed
on $[-1,1]$, \textcolor{black}{$a$ is the conservative force magnitude, $\gamma$ is the dissipative 
force coefficient, $k$ is the power index of the dissipative force envelop, $r_c$ 
is the cutoff distance of DPD interaction, $k_s$ is the
bond spring constant, and $r_{max}$ is the maximum bond extension distance,
respectively. $\left\langle \cdots \right\rangle$ and
$\sigma$ represents the magnitude of uncertainty for each parameter.}
Similar to the Newtonian DPD fluid systems, we choose the mean values as the
standard force parameters employed in previously published DPD simulations \cite{Groot_DPD_1997}. 
To investigate the numerical performance of the compressive sensing
method on different parameter spaces, we choose three parameter sets with
different variance values, as shown in Table \ref{tab:melt_params}.

In this study, we aim to study the shear rate dependent viscosity of 
the polymer melt system. In general, numerical evaluation of this property
is a non-trivial task. The shear viscosity of a fluid system is usually
determined either by the linear response theory with calculation of the stress
auto-correlation function terms in equilibrium state, or by modeling the steady
shear flow with the Lees-Edwards boundary conditions \cite{Lees_BC_1972}.
However, both approaches incorporate the calculation of the stress field, which
is extremely noisy in equilibrium or low shear rate regime. Therefore, a large
numerical error is anticipated in the regime where the thermal fluctuation
dominates over the ensemble average value. Moreover, unlike a simple Newtonian
fluid system, polymeric fluids typically exhibit non-Newtonian behavior with
shear rate dependent viscosity. As shear rate approaches zero, the normal stress
and viscosity value typically approach a constant \textit{low-shear-rate}
plateau, which is usually inaccessible due to rheometer limitations \cite{Larson98}.
In this study, we adopt the reverse Poiseuille flow rheometer briefly explained as below.

%-------------------------------------------------------------------------------
\begin{table}[!h]
\centering
\caption{DPD force parameters for non-Newtonian polymer melt system.
$\left<\star\right>$ and $\sigma_{\star}$ represent the average and
magnitude of the random variables for individual model parameters.  
The superscripts $l$, $m$ and $s$ represent the three parameter
sets with different variance values.}
\begin{tabular}{C{5em}*{6}{C{4.8em}}}
  \hline \hline
  & $a$ & $\gamma$ & $k$ & $r_c$ & $k_s$ & $r_{max}$ \\
  \hline
  $\left<\star\right>$ & $25.0$ & $8.0$ & $0.25$ & $1.0$ & $25.0$ & $1.0$ \\
  $\sigma_{\star}^{l}$ & $15.0$ & $4.0$ & $0.18$ & $0.06$ & $15.0$ & $0.06$ \\
  $\sigma_{\star}^{m}$ & $10.0$ & $2.0$ & $0.15$ & $0.05$ & $10.0$ & $0.05$ \\
  $\sigma_{\star}^{s}$ & $6.0$ & $1.2$ & $0.08$ & $0.03$ & $6.0$ & $0.03$ \\
  \hline \hline
\end{tabular}
\label{tab:melt_params}
\end{table}
%-------------------------------------------------------------------------------

\subsection{Shear viscosity rheometer}
\label{sec:vis_rheometer}
%-------------------------------------------------------------------------------
\begin{figure}[!h]
\centering
\includegraphics*[scale=0.5]{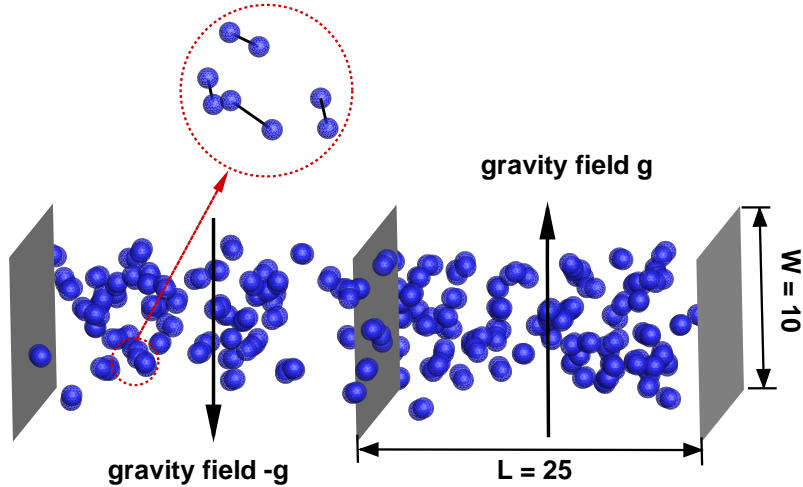}
\caption{Sketch of the simulation polymer melt system. $N_p = 15,000$ polymer
chains are placed in the simulation domain with size $50\times20\times10$ in DPD
reduced units (only $2\%$ of the simulation chain is visualized). Each polymer
chain consists of $N_b = 2$ DPD particles, represented by the particles in blue
color. Individual polymer chains are connected by the FENE bond potential, as
defined by Eq. \eqref{eq:fene}. \textcolor{black}{The grey borders 
refer to the boundaries where body force changes direction.}}
\label{fig:simulation_box}
\end{figure}
%-------------------------------------------------------------------------------

We compute the \textit{low-shear-rate} viscosity values using the
reverse Poiseuille flow rheometer. This approach was initially proposed for
computing the shear viscosity for Newtonian flow system. However, a later study
showed that this approach can be well extended to study the complex fluid system
with non-Newtonian behavior \cite{Fedosov_Polymer_2010}. This approach generates
more accurate results than the previous two methods as it replaces the expensive
stress calculation with the calculation of velocity profile. Here we briefly
review the method and we refer to \cite{BacLow05, Fedosov_Polymer_2010}
for details.

As shown in Figure \ref{fig:simulation_box}, we simulate $N_p = 15,000$ polymer
chains in a domain of $50 \times 20 \times 10$ DPD units with periodic boundary
conditions. The domain is divided into two regimes at center ($x=0$), where an
equal but opposite gravity force $f$ between $0.015$ and $0.05$ is applied on
each DPD particle in each half of the domain. At the cross-stream position $x$
and time $t$, we have
%-------------------------------------------------------------------------------
\begin{equation}
\rho \frac{\partial u}{\partial t} = \frac{\partial \tau_{xy}}{\partial x}-fn,
\label{eq:motion}
\end{equation}
%-------------------------------------------------------------------------------
where $n$ is the number density. Under steady state flow, the left hand side of
Eq. \eqref{eq:motion} vanishes and the shear stress $\tau_{xy}$ is linear across
the channel with maximum value $fnH/2$ across the virtual wall boundary, where
$H = 50$ is the length of the channel.

Having computed the shear stress profile, the shear-rate-dependent viscosity
$\eta(x)$ can be calculated by
%-------------------------------------------------------------------------------
\begin{equation}
\tau_{xy}(x) = \eta(x){\dot{\gamma}(x)},
\label{eq:rheol}
\end{equation}
%-------------------------------------------------------------------------------
where $\dot{\gamma}(x)$ is the shear rate across the channel. Therefore, the
calculation of shear viscosity is transformed into the calculation of the
velocity derivatives across the channel. For the Newtonian fluid system, the
shear viscosity across the channel is nearly constant and the steady flow
velocity maintains a parabolic shape. For the non-Newtonian
polymer melt system,
the shear viscosity is inhomogeneous across the channel. In particular, the
\textit{low-shear-rate} viscosity values are extracted from a flattened velocity
profile and hence a relatively larger numerical error is
anticipated. To further reduce the numerical error, we fit the velocity profile
of Newtonian fluid and polymer melt system using a $2$nd and $4$th-order
polynomial, respectively:
%-------------------------------------------------------------------------------
\begin{subequations}
\begin{align}
V(x) &= V_c \pm C_2 (x \pm H/4)^2, \label{eq:quadratic_fit} \\
V(x) &= V_c \pm C_2 (x \pm H/4)^2 \pm C_4 (x \pm H/4)^4,
\label{eq:fourth_fit}
\end{align}
\label{eq:vel_poly_fit}
\end{subequations}
%-------------------------------------------------------------------------------
\begin{figure}[!h]
\centering
\includegraphics[scale=0.5]{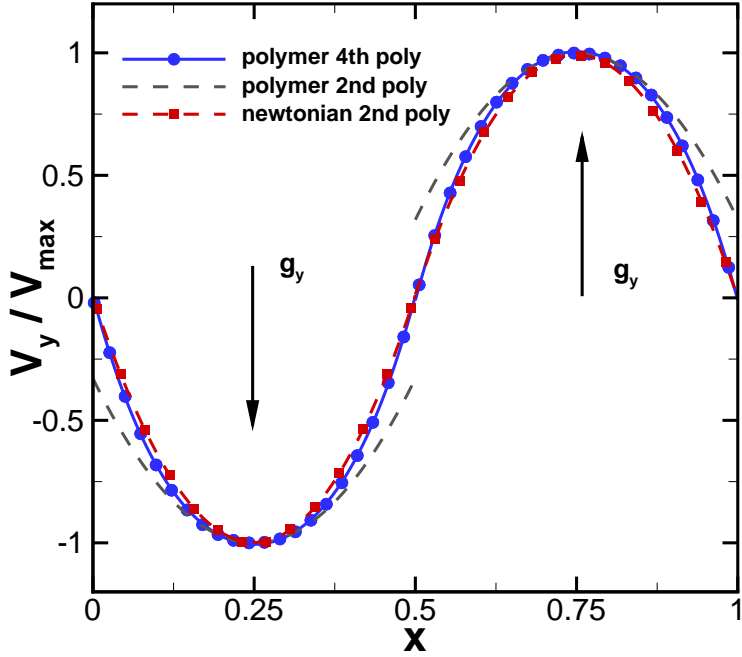}
\caption{Typical steady reverse Poiseuille flow velocity profiles of the
Newtonian and polymer melt system. The points represent the direct simulation
results. The solid line represents the 4th order polynomial fitting curve for
the polymer melt system. The dash lines represent the 2nd order polynomial
fitting curves for the polymer melt and Newtonian flow systems.}
\label{fig:vel_poly_fit}
\end{figure}
%-------------------------------------------------------------------------------
\begin{figure}[!h]
\centering
\includegraphics[scale=0.5]{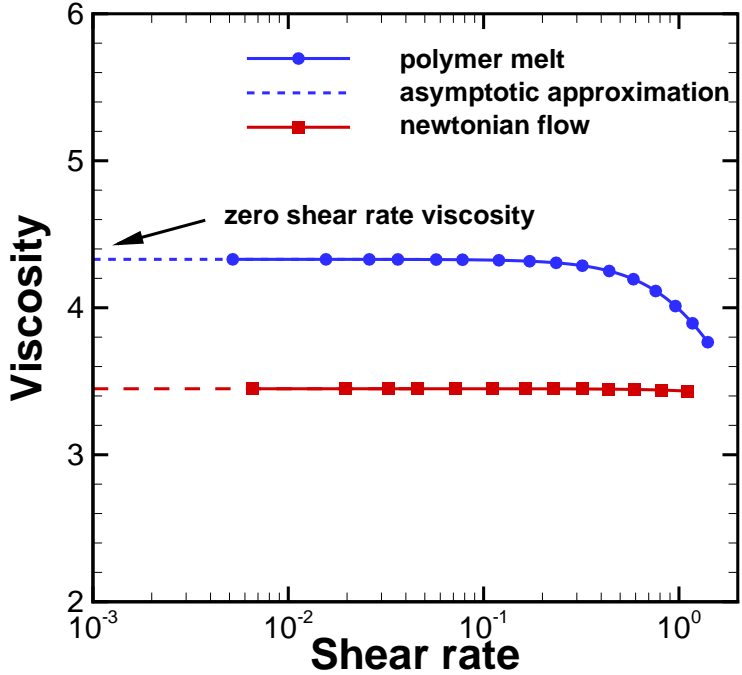}
\caption{\textit{Shear-rate-dependent} viscosity $\eta(\dot{\gamma})$ computed
from the simulated velocity profiles for the polymer melt given $a = 20.0$,
$\gamma = 8.0$, $k = 0.26$, $r_c = 1.0$, $k_s = 40.0$ and $r_{max} = 1.0$ and
Newtonian flow systems given $a = 40.0$, $\gamma = 4.5$ and $k = 0.2$. 
\textcolor{black}{The
\textit{zero-shear-rate} viscosity of the polymer melt system is computed by
taking the limit of asymptotic approximation (the dash line) 
$\displaystyle \lim_{\dot{\gamma} \to 0} \eta(\dot{\gamma}) \approx 4.35$ using Eq. (\ref{eq:vel_poly_fit}).}}
\label{fig:shear_rate_viscosity}
\end{figure}
%-------------------------------------------------------------------------------

For each parameter set, we conduct three independent reverse Poiseuille flow
simulations: $800,000$ steps are performed for each simulation and the velocity
profile sampling is taken during the last $700,000$ steps; a time step
$dt = 0.01$ is adopted for all the simulations. Figure \ref{fig:vel_poly_fit}
shows an example of the steady velocity distribution across the channel for both
the polymer melt and Newtonian fluid systems. The Newtonian fluid system
maintains a near constant viscosity value across the channel and the velocity
profile can be well fitted by Eq. (\ref{eq:quadratic_fit}). On the contrary, the
polymer melt system exhibits varying viscosity value across the channel, resulting
in a poor fitting by Eq. \eqref{eq:quadratic_fit}. Instead, high order terms in
Eq. \eqref{eq:fourth_fit} are essential to capture the non-Newtonian properties.
With the fitted velocity profile, the shear viscosity of the Newtonian fluid is
determined by $\frac{n f}{2 C_2}$ and the \textit{zero-shear-rate} viscosity is
computed by taking the limit $\eta_0 = \lim_{x \rightarrow \pm H/4} \eta(x)$,
where $\eta (x)$ is \textit{low-shear-rate} viscosity determined by the local
shear viscosity determined by Eq. \eqref{eq:rheol} and
Eq. \eqref{eq:vel_poly_fit}, as shown in Figure \ref{fig:shear_rate_viscosity}.
%-------------------------------------------------------------------------------

\subsection{Non-isothermal liquid water system}
\label{subsec:edpd_model}
The liquid water is employed as an example of non-isothermal system, where the diffusivity
and viscosity of liquid water at various temperatures ranging from $273~K$ to $373~K$ are set
as target properties. For water, the experimental data of its self-diffusivity $D$ at various 
temperature is taken from \cite{Holz00} and the kinematic viscosity $\eta$ is taken
from \cite{Bergman11}. In particular, at $T^*=300K$ the diffusivity of liquid water is
$2.41\times 10^{-9} m^2s^{-1}$ and the kinematic viscosity is $8.57\times 10^{-7} m^2s^{-1}$, 
hence the Schmidt number $\rm{Sc}=\eta / D = 355.60$. 

Dimensionless variables that are
suitable for interpretation of the results are introduced to carry out the
simulations. The temperature of reference $T^*=300K$ is used to scale the
temperature, thus the temperature ranging from $273K$ to $373K$ are
represented by the dimensionless temperature $T=0.91$ to $1.2433$.
To capture the correct Schmidt number at $T=1.0$ the parameters in eDPD system
are defined as
$n=4.0$, $a_{ij}=75k_BT/n$, $\gamma=4.5$, $\sigma^2=2k_BT\gamma$,
$r_c=1.58$ and $s(T)|_{T=1.0}=0.41$, which yields an eDPD fluid with diffusivity 
$D=1.87\times10^{-2}$, kinematic viscosity $\eta=6.62$ and corresponding 
Schmidt number is $Sc=354.01$ with respect to the experimental data $Sc=355.60$. 
Moreover, the heat capacity is $C_v=1.0\times10^5$ and $\kappa=315k_B\eta/2\pi\rho C_v r_c^5 Pr$ 
in which $Pr$ is the Prandtl number. More computational details can be found in \cite{LiTLBK14}.

Considering the effect of temperature on the dynamic properties of eDPD fluid, $s$ 
(the exponent of the weight function in the heat flux equations \eqref{equ:heat1} and \eqref{equ:heat3}) 
is defined as a function of the
temperature to reproduce the experimental data of liquid water over a range of $273~K$ to $373~K$. In the
present work, $s$ is set as the following polynomial:
\begin{equation}
\label{eq:edpd_exp}
s(T)=C_0+C_1(T-1) + C_2(T-1)^2 + C_3(T-1)^3 + C_4(T-1)^4,
\end{equation}
where $C_0$ is set to $0.41$ so that the Schmidt number matches the experimental data 
at $T=1.0$. The other coefficients $C_i, i=1,2,3,4$ will be
identified through Bayesian inference to get consistent transport properties
with the experimental data of liquid water, which will be described 
in detail in section \ref{sec:eDPD}.

%===============================================================================
%===============================================================================
%===============================================================================

%\section{Stochastic sparse modeling}
\section{Constructing the surrogate model}
\label{sec:gpc_l1}
%\textcolor{green}{Xiu, can you please add brief outline for this section?}
In this section, we introduce an efficient approach to build the surrogate
model of the target properties. We will briefly review the gPC approximation and
the compressive sensing {method}, then demonstrate the algorithm of combining
these two approaches to construct the surrogate model.

\subsection{gPC expansion for target property}
\label{subsec:pcm}
 Due to the linear relation between
$\bm\theta$ and $\bm\xi$ in this paper (e.g., see
Eq. \eqref{eq:para_uncertain}), we consider a gPC expansion depending on $\bm\xi$ instead.
Given a set of parameter $\bm\xi$, the output $X(\bx)$ obtained from the DPD
simulation is
%-------------------------------------------------------------------------------
\begin{equation}
X({\bm\xi}) = \mu({\bm\xi}) + \phi,
\end{equation}
%-------------------------------------------------------------------------------
where $\mu(\bx)$ is the value of target property, and $\phi$ represents the
intrinsic thermal noise discussed in the system. In order to compute
$\mu(\bm\xi)$, we simulate multiple replicas for the parameter set $\bx$. As the
number of replica increases, the average of $X(\bm\xi)$ converges to
$\mu(\bm\xi)$ as we assume $\phi$ to be a Gaussian random variable. In the
present work, we take three identical independent simulations for
each $\bx$ and consider the average $\bar X(\bx)$ as a good approximation of
$\mu(\bx)$. Following the same approach as in \cite{Najm_UQ_2012_a}, we verify that
in the cases we study, the magnitude of the fluctuation is $\mo(0.1\%)$ (or
smaller) of the target property. Hence, the error from the intrinsic
thermal noise is much less than
%from quantifying the parameter uncertainty;
the error of the surrogate model;
this will be validated in Section \ref{sec:result}.

Given $\bar{X}(\bm\xi)$, we employ gPC to represent the target property:
%-------------------------------------------------------------------------------
\begin{equation}\label{eq:gpc1}
\bar{X}(\bx)=\sum_{|\ba|=0}^{\infty} c_{\ba}\psi_{\ba}(\bx),
\end{equation}
%-------------------------------------------------------------------------------
where $\ba$ is a multi-index, $\bx=(\xi_1,\xi_2,\cdots,\xi_d)$ is a vector of
$d$ \textit{i.i.d.} random variables, $\psi_{\ba}$ is a set of orthonormal
polynomials associated with the probability measure $\nu$ of $\bx$ and $c_{\ba}$
is the coefficient. We truncate the expression \eqref{eq:gpc1} up to polynomial
order $P$, hence $\bar{X}$ is approximated as:
%-------------------------------------------------------------------------------
\begin{equation}\label{eq:gpc2}
\bar X(\bx)\approx\widetilde{X}(\bx)=\sum_{|\ba|=0}^{P}c_{\ba}\psi_{\ba}(\bx).
\end{equation}
%-------------------------------------------------------------------------------
We use $\bm c$ to denote the vector of the gPC coefficients, i.e.,
$\bm c = (c_{\ba_1}, c_{\ba_2}, \cdots)$. By sorting the indices $\ba_i$ we can
also write it as $\bm c = (c_0, c_1, \cdots)$, where $c_0$ is the coefficient of
the zero-th order gPC basis function.
The gPC representation $\widetilde{X}$ is constructed by computing the
coefficients $c_{\ba}$. This procedure can be accomplished by
using the probabilistic collocation method (PCM) e.g., tensor product
points or sparse grid points as mentioned in the introduction: after obtaining
$\bar X^1, \bar X^2,\cdots,\bar X^M$, which are values of $\bar X$ at sampling
points $\bx^1,\bx^2,\cdots,\bx^M$, we compute the gPC coefficients of $\bar X$ :
\begin{equation}
\label{eq:coef}
c_{\alpha} = \dfrac{\int \bar X(\bx)\psi_{\alpha}(\bx)\dif\nu(\bx)}
{\int \psi^2_{\alpha}(\bx)\dif\nu(\bx)},
\end{equation}
(where $\nu$ is the probability measure of $\bx$ and the integrals are
multi-variate integrals depending on the dimension of $\bx$) through
approximating the integral with quadrature rule, e.g.,
\begin{equation}
\int \bar X(\bx)\psi_{\alpha}(\bx)\dif\nu(\bx)\approx\sum_{i=1}^M \bar
X^i\psi_{\ba}(\bx^i) w^i.
\end{equation}
Here $w^i$ is the corresponding weight for $\bx^i$.

\subsection{Sparsity and compressive sensing}\label{sec:algo}
For systems with relatively high dimensions, e.g., the polymer melt
model with $d=6$, the method to compute the gPC coefficients described in
Section \ref{subsec:pcm} may not be efficient in that in order to increase
the accuracy, the increment of the number of simulations is fixed and this
number can be very large ($\mathcal{O}(10^2)\sim\mathcal{O}(10^3)$ or larger),
hence, these methods can be prohibitive for DPD simulations which are usually
very costly. Another drawback of this method is that it requires results from
all the sampling points. If the DPD code fails at some sampling points, the
method fails, although we may employ other approaches to obtain a less accurate
result. To overcome the above difficulties, we compute the gPC expansion by
applying the compressive sensing method as post processing for the Monte Carlo method
as discussed below.

We first generate $M$ random samples of the parameter sets $\bx^j, j=1,\cdots, M$
based on the distribution of the random variables. Notice that $\bx^j$ is a
random vector with \textit{i.i.d} entries. In this paper, we generate
$M$ such random vectors based on the uniform distribution $\mathcal{U}[-1,1]$.
Then we input these vectors in the DPD code to obtain $M$ outputs
$\bar{\bm X}=(\bar X^1,\bar X^2,\cdots,\bar X^M)$, respectively. Based on
\eqref{eq:gpc2}, we
obtain the following linear system:
%-------------------------------------------------------------------------------
\[\begin{pmatrix}
\psi_{\bm\alpha_1}(\bm\xi^1) & \psi_{\bm\alpha_2}(\bm\xi^1) & \cdots \\
\psi_{\bm\alpha_1}(\bm\xi^2) & \psi_{\bm\alpha_2}(\bm\xi^2) & \cdots \\
& \vdots &
\end{pmatrix}
\begin{pmatrix}
c_{\bm\alpha_1} \\  c_{\bm\alpha_2} \\ \vdots
\end{pmatrix}=
\begin{pmatrix}
\bar X(\bm\xi^1) \\  \bar X(\bm\xi^2) \\ \vdots
\end{pmatrix} + \bm\ve,\]
%-------------------------------------------------------------------------------
or equivalently,
%-------------------------------------------------------------------------------
\begin{equation}\label{eq:linear_sys}
\newtensor\Psi \bm c = \bar{\bm X} + \bm\ve,
\end{equation}
%-------------------------------------------------------------------------------
where $\newtensor\Psi$ is the ``measurement matrix" with entries
$\newtensor\Psi_{i,j}=\psi_{\ba_j}(\bx^i)$, $\bm c$ is the vector of the gPC
coefficients, $\bmx$ is the vector consisting of the outputs and $\bm\ve$ is
related to the truncation error. Here $\newtensor\Psi_{\ba_j}$ is the tensor product of
normalized Legendre polynomials since $\bx$ is uniformly distributed. According
to the compressive sensing theory, when $\bm c$ is sparse and $\newtensor\Psi$
satisfies certain conditions, we can recover it from Eq. \eqref{eq:linear_sys} by
solving the following $\ell_h$ minimization problem \cite{CandesRT06, BrucksteinDE09}:
%-------------------------------------------------------------------------------
\begin{equation}\label{eq:p1d}
(P_{h,\delta}):\qquad\min_c\Vert\bm c\Vert_h\quad\mbox{subject to}\quad
\Vert\newtensor\Psi\bm c-\bmx\Vert_2\leq\delta,
\end{equation}
%-------------------------------------------------------------------------------
where $h=1$ or $0$, $\delta=\Vert\bm\ve\Vert_2, \Vert\bm c\Vert_1=\sum |c_i|$
and $\Vert\bm c\Vert_0=$ number of nonzero entries of $\bm c$.

In this paper we employ $\ell_0$ minimization, i.e., the results
of $(P_{0,\delta})$. To solve this minimization problem, we employ the greedy
algorithm \textit{orthogonal matching pursuit} (OMP) presented in
\textbf{Algorithm 1} \cite{BrucksteinDE09}. 

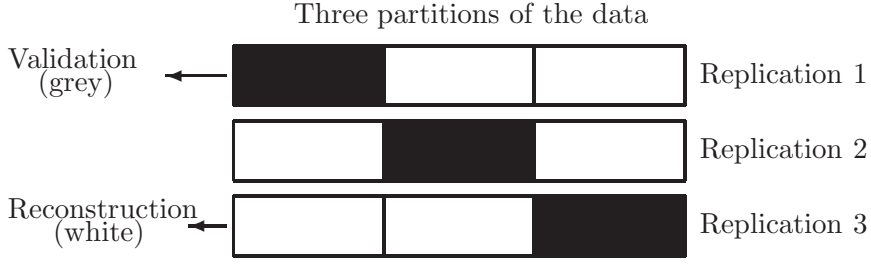
\begin{figure}[!h]
\centering
\setlength{\unitlength}{1mm}
\begin{picture}(100,35)
\linethickness{1pt}
\put(30,0){\line(1,0){60}} \put(30,8){\line(1,0){60}}
\put(30,0){\line(0,1){8}} \put(50,0){\line(0,1){8}}
\put(70,0){\line(0,1){8}} \put(90,0){\line(0,1){8}}
\put(30,10){\line(1,0){60}} \put(30,18){\line(1,0){60}}
\put(30,10){\line(0,1){8}} \put(50,10){\line(0,1){8}}
\put(70,10){\line(0,1){8}} \put(90,10){\line(0,1){8}}
\put(30,20){\line(1,0){60}} \put(30,28){\line(1,0){60}}
\put(30,20){\line(0,1){8}} \put(50,20){\line(0,1){8}}
\put(70,20){\line(0,1){8}} \put(90,20){\line(0,1){8}}
\put(92,3){Replication 3} \put(92,13){Replication 2} \put(92,23){Replication 1}
\linethickness{11pt}
\put(30,22){\line(1,0){20}} \put(30,26){\line(1,0){20}}
\put(50,12){\line(1,0){20}} \put(50,16){\line(1,0){20}}
\put(70,2){\line(1,0){20}} \put(70,6){\line(1,0){20}}
\put(38,31){Three partitions of the data}
\linethickness{1pt}
\thicklines
\put(29,24){\vector(-1,0){8}}\put(0,25){Validation}
\put(4,22){(grey)}
\put(29,4){\vector(-1,0){5}}\put(0,5){Reconstruction}
\put(6,2){(white)}
\end{picture}
\caption{Demonstration of cross-validation. The data is partitioned into three
parts. For each replication, the white parts are used in reconstruction and the
black part is used for validation.}
\label{fig:cross_valid}
\end{figure}

The threshold $\epsilon_0$ in Step 6 is set as $\delta$ in Eq. \eqref{eq:p1d} 
obtained by the cross-validation method. 
Typically, the magnitude of the truncation $\delta$ is not known
\textit{a priori}, and we use a cross-validation method to estimate it. We first
divide the $M$ available output samples into $M_r$ reconstruction samples (white
parts in Fig. \ref{fig:cross_valid}) and $M_v$ validation samples (black part in
Fig. \ref{fig:cross_valid}) such that $M=M_r+M_v$. Then repeat the OMP
method on the reconstruction samples with multiple choices of
truncation error tolerance $\delta_r$. This is accomplished by setting
$\delta$ in Eq. \eqref{eq:p1d} as different $\delta_r$. In
this paper we test $11$ different $\delta_r$ ranging from $10^{-4}$ to
$10^{-2}$ with constant ratio. For each $\delta_r$, we solve $(P_{0,\delta_r})$
to obtain gPC coefficient $\bm c_r$ and estimate the error
$\delta_v=\Vert\newtensor\Psi_v\bm c_r-\bm u_v\Vert_2$ with the validation data.
Next, we repeat the above cross-validation algorithm for multiple replications
with different selection of reconstruction and validation samples. Finally, the
estimate of $\delta=\sqrt{M/M_r}\hat\delta_r$ is based on the values of
$\hat\delta_r$ for which the average of the corresponding validation errors
$\delta_v$, over all replications of the validation samples, is minimum. In
this paper we partition the data into three parts, i.e., we set
$M_r\approx 2M/3$ and perform the cross-validation for three replications.
More details can be found in \cite{DoostanO11}.

The algorithm of the entire procedure is presented in \textbf{Algorithm 2}. In
this paper, the OMP method is achieved by \verb SparseLab  \cite{DonohoDST}.
Finally, since $M$ different samples of $\bx^j$ lead to $M$ different
$\bar X^j, j=1,2,\cdots, M$, we use ``number of samples" to refer to $M$ in
Section \ref{sec:result}.

%-------------------------------------------------------------------------------
\begin{algorithm}
\hrule
\caption{Approximate the solution of $(P_0)$: $\min_{\bm c}\Vert\bm c\Vert_0$
  subject to {$\Vert\newtensor\Psi\bm c - \bm b\Vert_2\leq \delta$}
  by the OMP method \cite{BrucksteinDE09}.}\vspace{5pt}
  \hrule \vspace{5pt}
\begin{algorithmic}[1]
\State Initialize $k=0$, $\bm c^0=0$, residual {\color{black}$\bm r^0=\bm
\bar{\bm X}-\newtensor\Psi\bm c^0=\bm b$}, solution support
$\mathcal{S}^0=\text{Support}\{\bm c^0\}=\emptyset$.
\State $k = k+1$, compute
$\epsilon(j)=\min_{z_j}\Vert \newtensor\Psi_jz_j-\bm r^{k-1} \Vert_2^2$ for all
$j$ with $z_j^*=\newtensor\Psi_j^T\bm r^{k-1}/\Vert\newtensor\Psi_j\Vert_2^2$.
\State Find a minimizer $j_0$ for $\epsilon(j):\forall j\notin S^{k-1}$,
$\epsilon(j_0)<\epsilon(j)$  and update
$\mathcal{S}^k=\mathcal{S}^{k-1}\cup\{j_0\}$.
\State Compute $\bm c^k$, the minimizer of
$\Vert\newtensor\Psi\bm c-\bm b\Vert_2^2$ subject to
$\text{Support}\{\bm c\}=\mathcal{S}^k$.
\State Update the residual: {\color{black}$\bm r^k=\bar{\bm X}-\newtensor\Psi\bm c^k$}.
\State Stop the iteration if $\Vert\bm r^k\Vert_2<\delta$, where
$\epsilon_0$ is the stopping threshold. Otherwise go to step 2.
\end{algorithmic}
\hrule
\label{alg:omp}
\end{algorithm}
%-------------------------------------------------------------------------------

%-------------------------------------------------------------------------------
\begin{algorithm}
\hrule
\caption{Compressive sensing method to obtain gPC surrogate model for
  DPD.}\vspace{5pt}
\hrule\vspace{5pt}
\begin{algorithmic}[1]
\State Generate $M$ sampling points $\bx^1, \bx^2, \cdots, \bx^M$ based on the
distribution of $\bx$.
\State Run the DPD code with inputs
$\bx^1, \bx^2, \cdots, \bx^M$ to obtain $M$ outputs
$\bar X^1, \bar X^2, \cdots, \bar X^M$, respectively. Denote
$\bar{\bm X}=(\bar X^1,\bar X^2,\cdots,\bar X^M)$ and it is the ``observation"
in $(P_{0,\delta})$.
\State Construct the $M\times N$ ``measurement matrix" $\newtensor\Psi$
as $\newtensor\Psi_{i,j}=\psi_{\ba_j}(\bx^i)$, where $\psi_{\ba_j}$ are the basis
functions, $N$ is the total number of basis functions depending on $P$ in \eqref{eq:gpc2}.
\State Set the tolerance $\delta$ in $(P_{0,\delta})$ by employing
cross-validation method. 
\State Solve the  $\ell_0$ minimization problem
\begin{equation*}
\bm c=\arg\min\Vert \bm c\Vert_0\quad\mbox{subject to}
\quad \Vert\newtensor\Psi\bm c-\bm u\Vert_2\leq\delta.\end{equation*}
\end{algorithmic}
\hrule
\label{alg:omp_dpd}
\end{algorithm}
%-------------------------------------------------------------------------------
%
%\subsection{Cross-validation}
%\label{subsec:cross_valid}
%-------------------------------------------------------------------------------

%===============================================================================
%===============================================================================
%===============================================================================

\section{Numerical Results}
\label{sec:result}
In this section we present the numerical results using the new framework
we developed. We first
construct the surrogate models of the polymer melt and non-isothermal liquid water systems
with respect to 6-dimensional and 4-dimensional stochastic parameter space, respectively. 
We then compare our results with the ones obtained from the sparse grid method 
\cite{XiuH05} (a standard collocation method in UQ studies.)
Finally, with the surrogate models of the various target properties, we 
can infer the force parameters given a specific set of target property values observed
at the macroscopic level, which enables us to optimize the parameter sets to
reproduce the target property values, and/or 
identify possible parameter degeneracy within the present mesoscopic models.

As we need to construct an accurate surrogate model to approximate the response
surface, we check the relative $L_2$ error:
\begin{equation}
\label{eq:l2_err0}
\ve_0 = \sqrt{\dfrac{\int |\mu(\bx) - \tilde X(\bx)|^2\dif\nu(\bx)}
            {\int |\mu(\bx)|^2\dif\nu(\bx)}},
\end{equation}
where $\mu$ is the target property and $\tilde X$ is the gPC expansion of $\bar X$
(see Eq. \eqref{eq:gpc2}). Since we approximate $\mu$ with $\bar X$, we check
the following error instead:
\begin{equation}
\label{eq:l2_err}
\ve = \sqrt{\dfrac{\int |\bar X(\bx) - \tilde X(\bx)|^2\dif\nu(\bx)}
            {\int |\bar X(\bx)|^2\dif\nu(\bx)}}.
\end{equation}
As mentioned in Section \ref{subsec:pcm}, in our numerical examples, the thermal
noise is about $10^{-3}$ (or smaller) of the target property, i.e.,
$|\mu-\bar X|/|\mu|\sim \mo(10^{-3})$. Take the shear-rate viscosity for
example, the relative error of approximating $\bar X$ by $\tilde X$ is
$\mo(10^{-2})$ (as we will see in this section), namely,
$|\bar X-\tilde X|/|\bar X|\sim \mo(10^{-2})$, which reflects that
$|\mu-\tilde X|/|\mu|\sim\mo(10^{-2})$. Hence, $\ve$ is a good approximation of
$\ve_0$.
The integrals in Eq. \eqref{eq:l2_err} are approximated by quadrature rules
based on the tensor product or by the sparse grid method. The error
%by these methods
of numerical integration
is far below the magnitude of the thermal noise as we use high order
quadrature points or a high level sparse grid method to approximate the integral.

%\textcolor{green}{Xiu, can you please add the remark about noise as we discussed?}
\begin{rem}
In this work, the thermal noise in the system is relatively small, hence
the approximation of the target property through a single set of gPC expansion
is applicable, and the comparison Eq. \eqref{eq:l2_err} is reasonable. For
systems with large thermal noise,
%we may need to include another set of gPC expansion to present the thermal noise.
we need to develop a more advanced technique, which we plan to do in future work.
To this end, we can employ the ANOVA method to separate out explicitly 
the intrinsic thermal fluctuations from the stochasticity introduced to the 
parametric uncertainty. This was accomplished in \cite{Maitre_Knio_2015} 
for a model problem.
\end{rem}

\subsection{Polymer melt system}
\label{subsec:poly_melt}
First we consider a polymer melt system with 6-dimensional
parameter space and study three different cases with different parameter sets
summarized as below:
\begin{itemize}  
\item {\em Mesoscopic system}: polymer melt model defined by Eq. (\ref{eq:DPD_eq2})
and Eq. (\ref{eq:fene}) with number density $n = 3.0$ and $k_BT = 1.0$.
\item {\em Model parameters of uncertainty}: $a(\xi_1)$, $\gamma(\xi_2)$, $k(\xi_3)$,
$r_c(\xi_4)$, $k_s(\xi_5)$ and $r_{max}(\xi_6)$ defined by Eq. (\ref{eq:para_uncertain})
and Tab. (\ref{tab:melt_params}).
\item {\em Surrogate model of target properties}: zero-shear-rate viscosity. 
\end{itemize}

We use the compressive sensing based method to construct a third-order gPC expansion as the
surrogate response surface. The total number of basis functions is $84$, i.e.,
the measurement matrix $\newtensor\Psi$ has $N=84$ columns. In this example,
the $L_2$ error defined in Eq.\eqref{eq:l2_err} is computed by approximating the
integrals with the level $4$ sparse grid method. The sparse grid
method mentioned in this section is based on Clenshaw-Curtis abscissas.

\begin{figure}[!h]
\centering
\subfigure[]{
\includegraphics[width=3in]{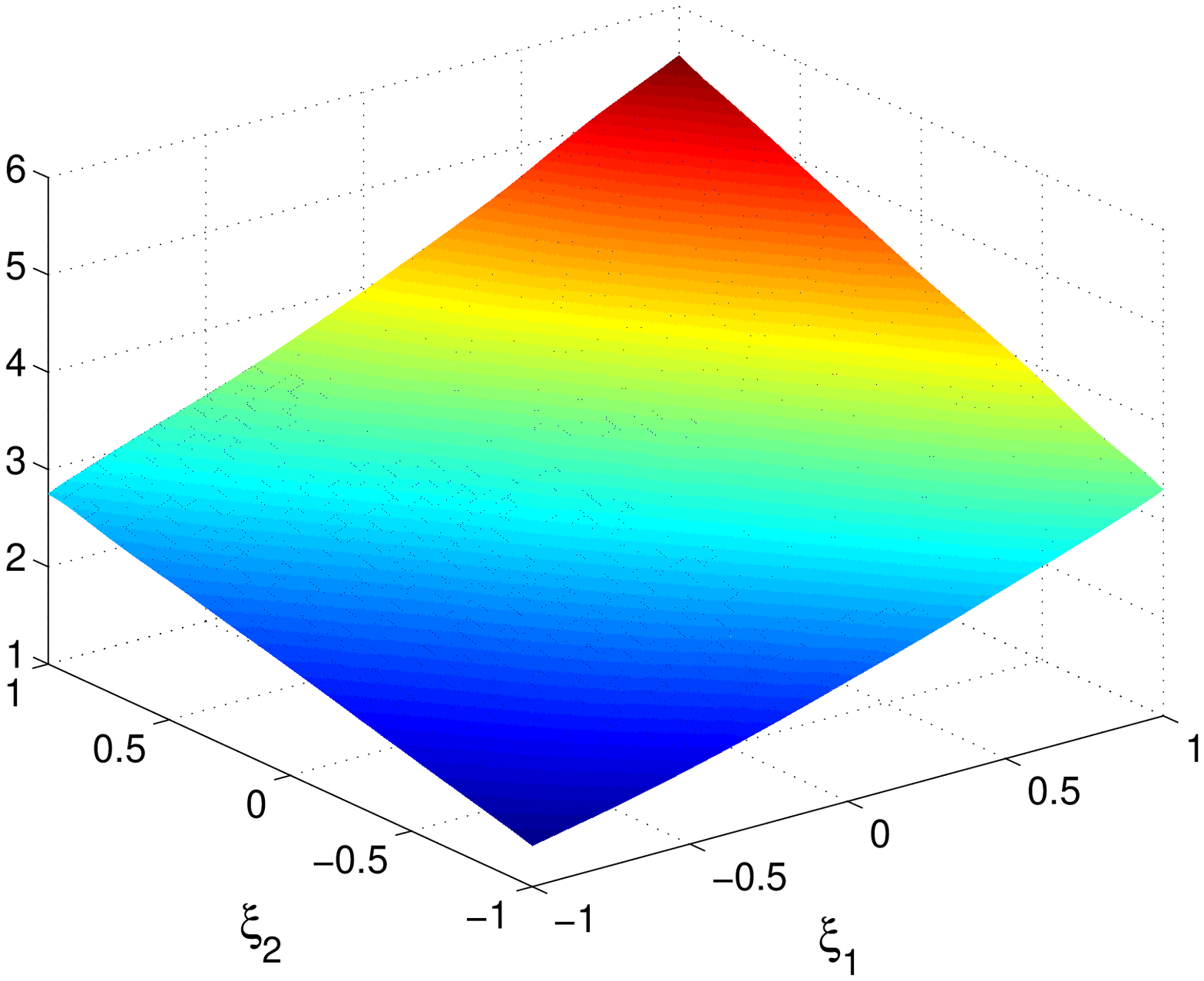}}
\subfigure[]{
\includegraphics[width=3in]{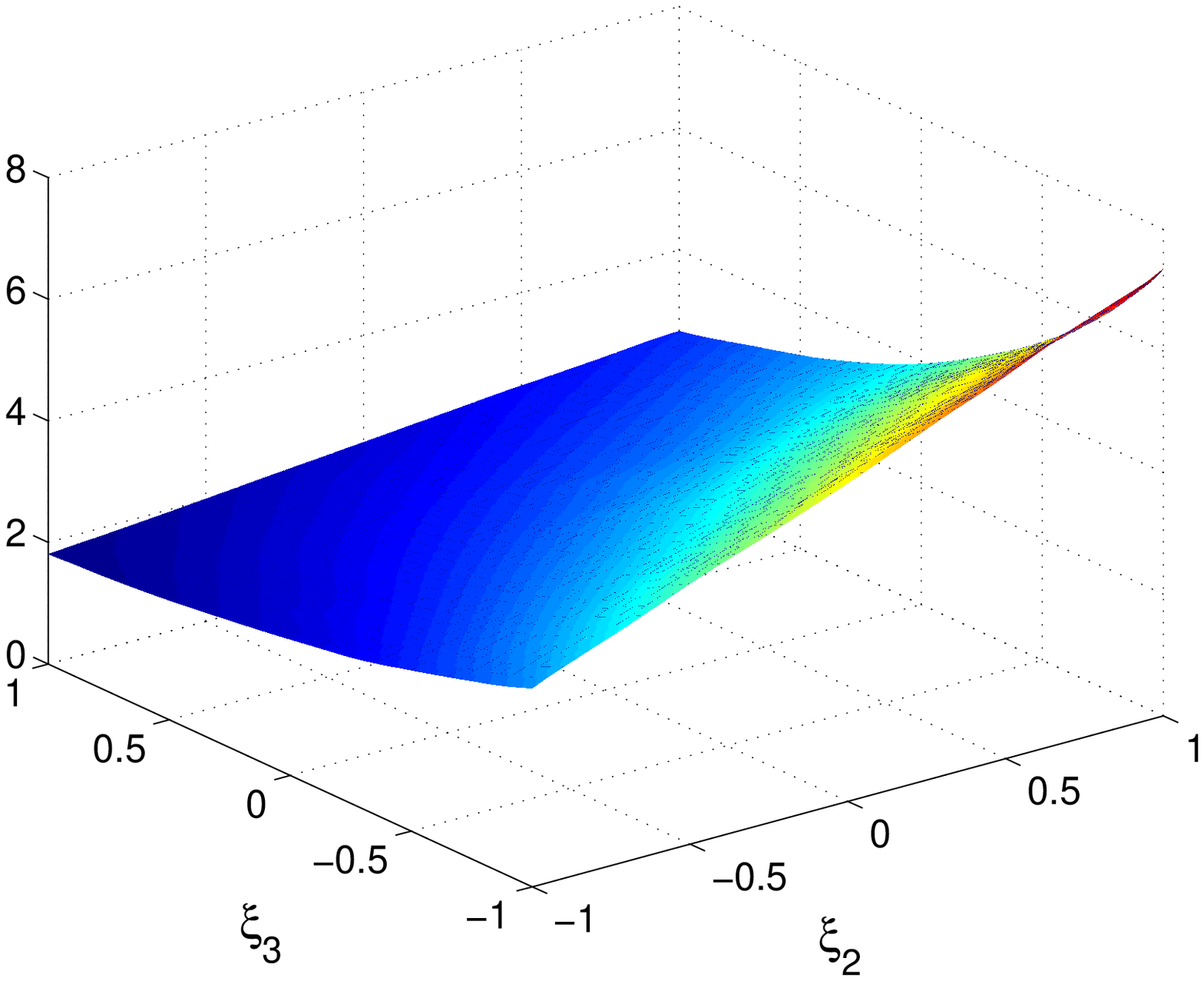}}
\subfigure[]{
\includegraphics[width=3in]{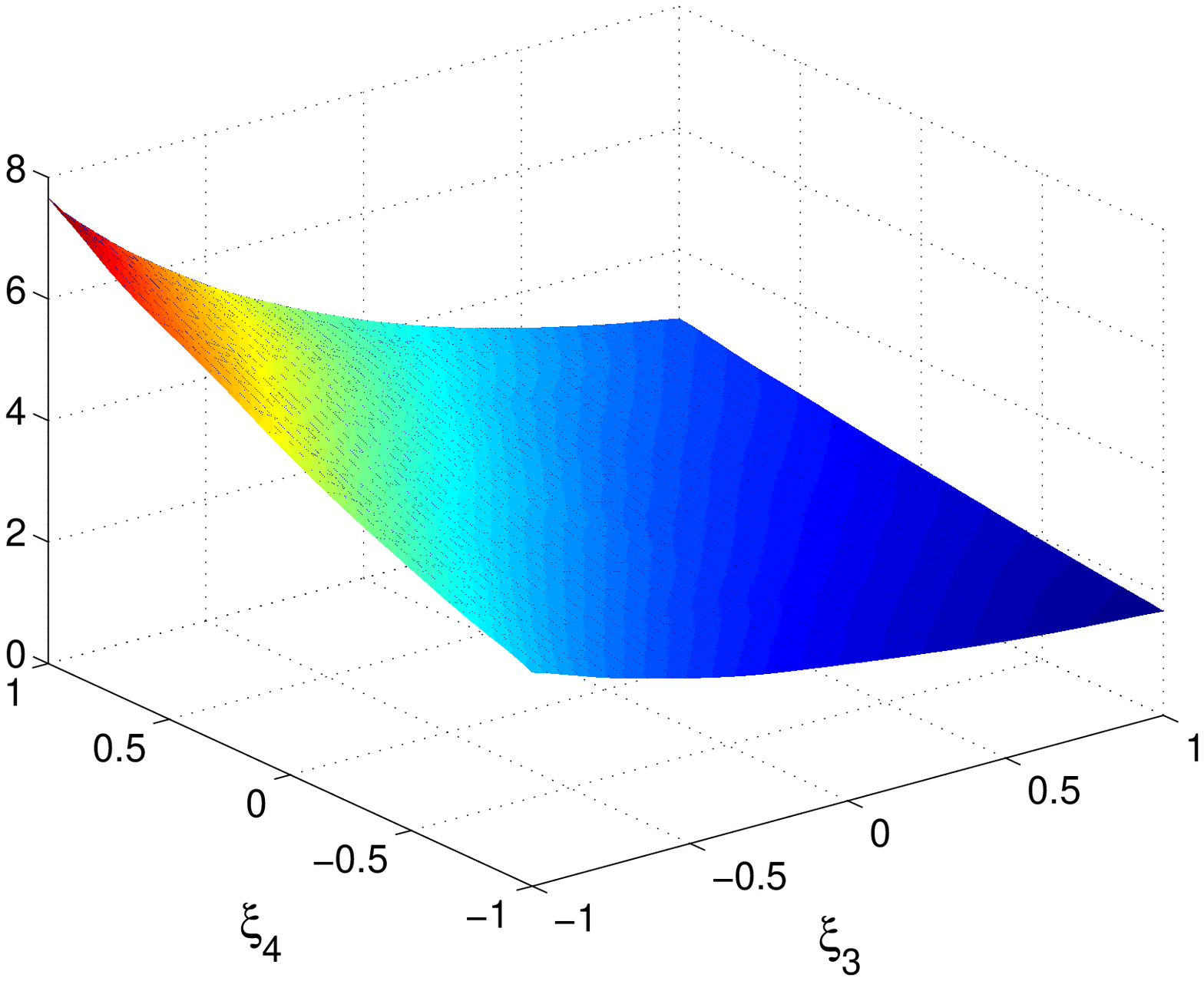}}
\subfigure[]{
\includegraphics[width=3in]{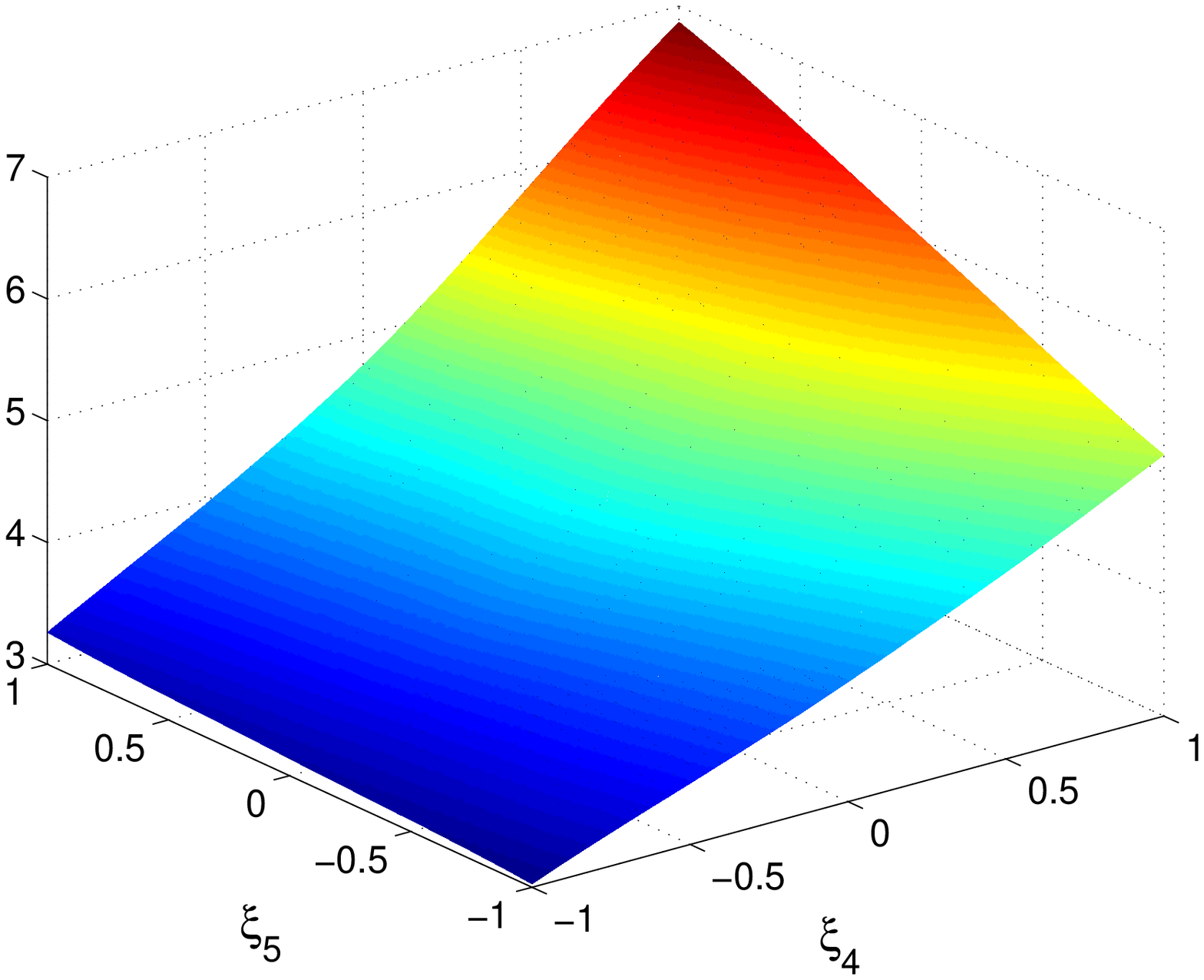}}
\caption{Response surface of the shear viscosity by fixing
the value of the rest entries of $\bx$. (a) $\bx_3 = \bx_4 = \bx_5 = \bx_6 = 0$,
(b) $\bx_1 = \bx_4 = \bx_5 = \bx_6 = 0$, (c) $\bx_3 = \bx_4 = \bx_5 = \bx_6 = 0$,
and (d) $\bx_1 = 1.0, \bx_2 = 0.0, \bx_3 = 0.0, \bx_6 = -1.0$.}
\label{fig:response}
\end{figure}

Since it is impossible to visualize a 7D hyper-plane, we present the response
surface of the shear viscosity by fixing part of the entries
of $\bx$ in Figure \ref{fig:response} to help understand the sensitivity of the
parameters.  These results are based on the simulation results of the
level 4 sparse grid method. For example, in
Figure \ref{fig:response}(a), we set $\xi_3=\xi_4=\xi_5=\xi_6=0$ to plot the
response surface with respect to $\xi_1$ and $\xi_2$.
The nearly planar response surface indicates that the momentum
transport is proportional to the magnitude of
the repulsive and dissipative force interactions. On the other hand, the curved
response surfaces in Figure \ref{fig:response}(b) and Figure \ref{fig:response}(c)
indicate the nonlinear dependence of the shear viscosity on the dissipative
interaction profile and the force interaction cut-off range. Moreover, the
response surface in Figure \ref{fig:response}(d)
indicates that the shear viscosity of the polymer melt system further depends on the
elasticity of the polymer model. To systematically study the
dependence of the shear viscosity on individual parameters, we compute the gPC
expansion using the OMP method introduced previously with fewer samples and use
the results by the level-4 sparse grid method point as reference solution. Similar
procedures are followed for different physical properties.
%-------------------------------------------------------------------------------
%-------------------------------------------------------------------------------

\begin{figure}[!h]
\centering
\includegraphics[width=3.5in]{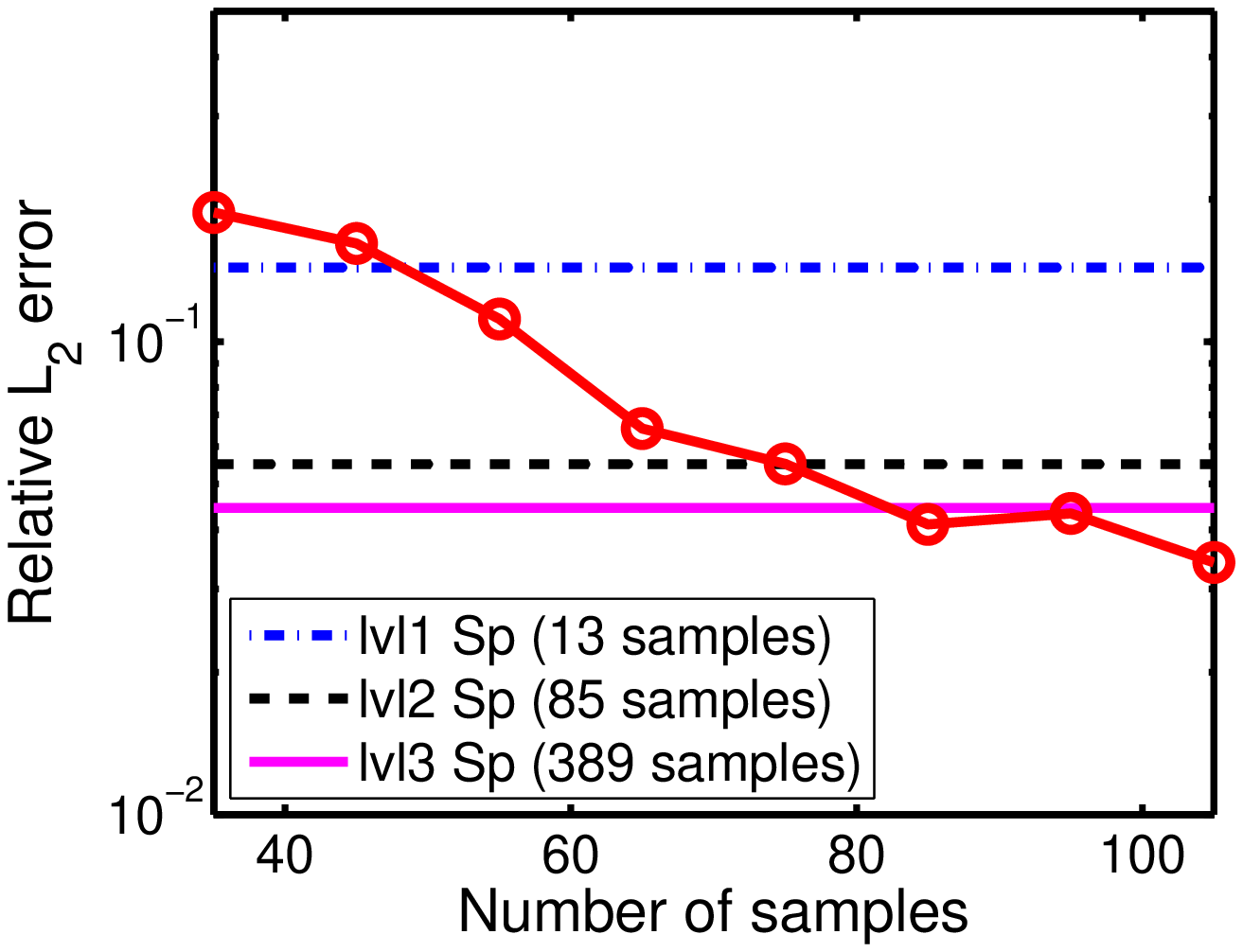}
\caption{$L_2$ error of the \textit{zero-shear-rate} viscosity of the polymer melt
model with parameter sets $\sigma_*^l$ by our method
  (``$\bigcirc$") is presented. Results of
sparse grid method (``Sp") of level 1, 2 and 3 are presented for comparisons.}
\label{fig:poly_large_l2_err}
\end{figure}

Figure \ref{fig:poly_large_l2_err} presents the $L_2$ errors with the parameter set
$\sigma_*^l$, i.e., the parameter set with the largest perturbation around
the mean, hence the response surface is more complicated than the parameter sets
$\sigma_*^m$ and $\sigma_*^s$. In this figure, when the number
of samples is larger than $75$ we also include $46$ gPC basis functions of the
fourth-order polynomials. We observe that the error of our method is decreasing
as more simulation results become available.
For example, if only $65$ simulations are affordable, our method can
provide a result with accuracy about the same level of that by the level-2 sparse
grid method. If $75$ or more samples are available, the accuracy of our method
exceeds that of the level-2 sparse grid method, which requires $85$ samples.
When $95$ or more samples are available, our method yields
more accurate results than the level-3 sparse grid method. As a comparison, if we
employ the level-3 sparse grid method, we need $304$ additional simulation results
given the $85$ simulation results for the level-2 sparse grid method. Hence,
our method is much more flexible than the standard sparse grid method.
%-------------------------------------------------------------------------------
%-------------------------------------------------------------------------------

\begin{figure}[!h]
\centering
\subfigure[$L_2$ error with $\sigma_*^m$]{
\includegraphics[width=3in]{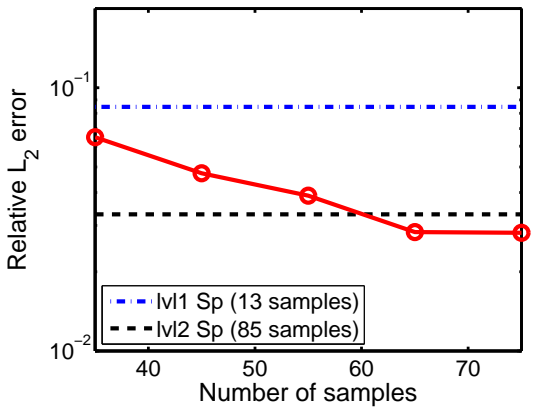}}
\subfigure[$L_2$ error with $\sigma_*^s$]{
\includegraphics[width=3in]{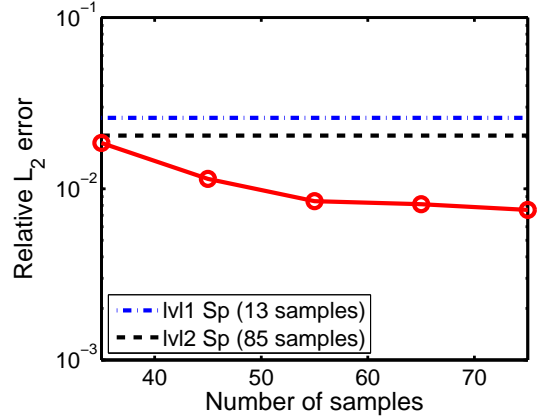}}
\caption{$L_2$ error of the \textit{zero-shear-rate} viscosity of the polymer melt
model with parameter sets $\sigma_*^m$ (left) and $\sigma_*^s$ (right)
  by our
method (``$\bigcirc$") are presented. Results of sparse grid method (``Sp") of
level 1 and level 2 are presented for comparisons.}
\label{fig:poly_medium_small_l2_err}
\end{figure}

Figure \ref{fig:poly_medium_small_l2_err} presents the same error analysis for the
parameter sets $\sigma_*^m$ and $\sigma_*^s$ and we only employ third-order gPC
expansions. It also illustrates the accuracy
and flexibility
of our method as Figure \ref{fig:poly_large_l2_err} does
since the error decreases as the number of samples increases,
and there is no restriction on the increment of the number of samples.
Moreover, by comparing Figure \ref{fig:poly_large_l2_err} and
Figure \ref{fig:poly_medium_small_l2_err}, we notice that in this model a smaller
perturbation in the parameters yields smaller difference between the results by
level-1 and level-2 sparse grid method. This is because the higher order basis
in the gPC expansion makes smaller contributions, which in turn implies that the
vector of the gPC coefficients is more sparse given a fixed
number of basis functions $N$. Therefore, we observe that with $55$ samples, our method
exceeds the level-2 sparse grid method for the parameter set $\sigma_*^s$ while
this is not the case for the other two parameter sets. This comparison implies
that the sparser the system is the more efficient our method is.
On the other hand, in order to increase the accuracy of the gPC
expansion, we may include more basis functions, which helps to reduce the
truncation error and also helps to increase the sparsity as long as the higher
order basis functions make small contributions. However, we cannot arbitrarily
increase the number of basis function since the compressive sensing algorithm
will fail if $N$ is too large given fixed number of samples. In this paper, we
keep the number of basis $M$ to be larger than $0.4N$.
%-------------------------------------------------------------------------------
%-------------------------------------------------------------------------------

%===============================================================================

\subsection{Inference of parameters in the model}

In this section, we investigate the parameter inference for the polymer melt systems and
the non-isothermal liquid water system. For polymer melt system, 
We first study a system with only $3$ varying force parameters (we fix the other
$3$ model parameters) using three target properties. Then we study a system with
$6$ varying model parameters using $6$ to $12$ target properties. For non-isothermal liquid water
system, we infer the $4$ model parameters using $6$ target properties. For reading clarity,
we summarize the setups of the mesoscopic systems at the beginning of each section.

\subsubsection{3D polymer melt model}
The setup of 3D polymer melt model is summarized as follows:
\begin{itemize}  
\item {\em Mesoscopic system}: polymer melt model defined by Eq. (\ref{eq:DPD_eq2})
and Eq. (\ref{eq:fene}) with
number density $n = 3.0$, $N_b = 4$ and $k_BT = 1.0$.
\item {\em Target properties for parameter inference}: \textit{zero-shear-rate} 
viscosity ($\eta$), the average value of radius of gyration ($R_g$), and the pressure ($P$)
with values specified by Eq. (\ref{eq:poly3Dtarget}).
\item {\em Model parameters of uncertainty}: $a(\xi_1)$, 
$k_s(\xi_5)$ and $r_{max}(\xi_6)$ with parameter confidence ranges specified
by Eq. (\ref{eq:para_uncertain2}).
\end{itemize}

Similar to the shear viscosity of the polymer melt system discussed in
Section \ref{subsec:poly_melt}, we can
use the OMP method to construct the response surfaces of various properties of
the systems, from which we are able to infer the different force parameters of the
mesoscopic model. As a simple demonstration, we consider a polymer melt system
with $N_b = 4$, $n = 3.0$ by fixing $(\gamma, k, r_c) = (8.0, 0.25, 1.1)$ while
the parameter space $(a, k_s, r_{max})$ is defined by
\begin{equation}
\begin{split}
 a(\xi_1) &= 25.0 + \sigma_{a}\xi_1,\\
 k_s(\xi_5) &= 60.0 + \sigma_{k_s}\xi_5,\\
 r_{max}(\xi_6) &= 0.8 + \sigma_{r_{max}}\xi_6,
\end{split}
\label{eq:para_uncertain2}
\end{equation}
where $(\sigma_{a}, \sigma_{k_s}, \sigma_{r_{max}}) = (15.0, 30.0, 0.2)$ and
$\bx = (\xi_1,\xi_5,\xi_6)$ are \textit{i.i.d} uniform random
variables on $[-1,1]$. In this test we construct a sixth-order gPC expansion with
Legendre polynomials as the surrogate model based on $54$ samples of DPD simulations.

For the system defined above, we target three bulk properties:
the \textit{zero-shear-rate} viscosity ($\eta$), the average value of radius of
gyration ($R_g$), and the pressure ($P$). Here $R_g$ of an individual polymer is defined
by
\begin{equation}
R_g^2 = \sum_{i=1}^{N_b} (\bm r_i - \bm r_c)^2/N_b
\end{equation}
where $\bm r_i$ and $\bm r_c$ represent the position of individual
bead $i$ and the center of mass, respectively.

\begin{figure}[!h]
\centering
\subfigure[$a$]{
\includegraphics[width=2in]{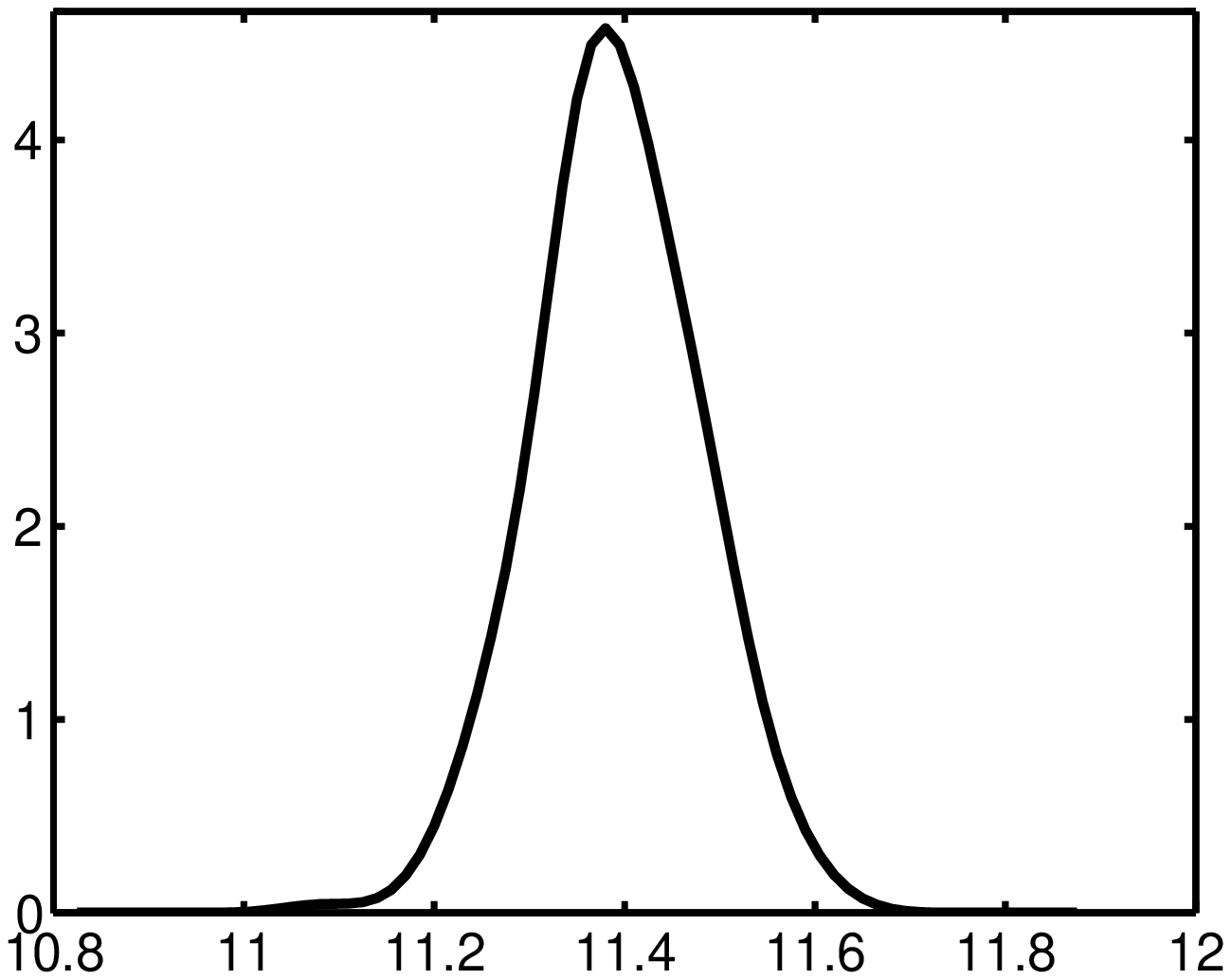}}
\subfigure[$k_s$]{
\includegraphics[width=2in]{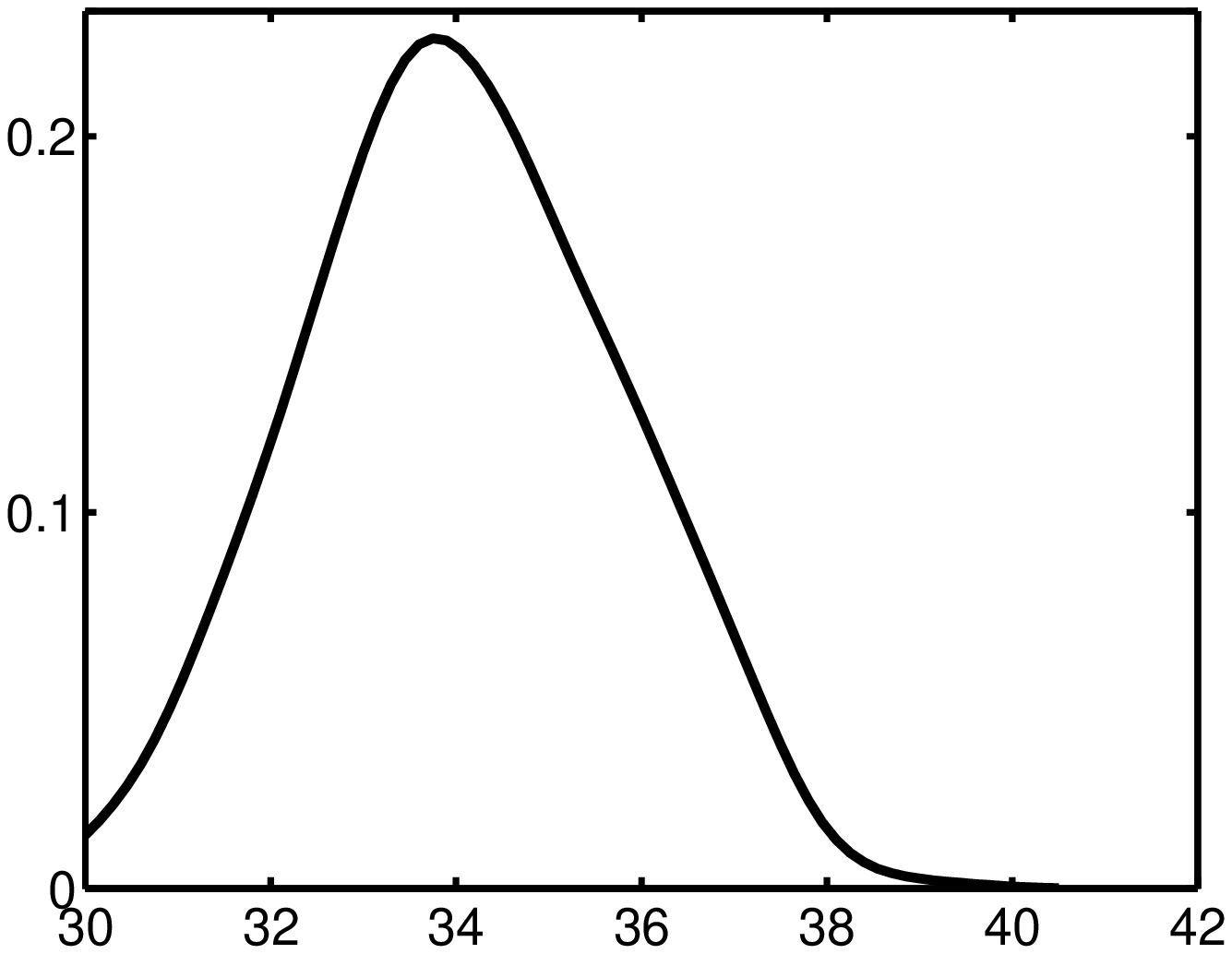}}
\subfigure[$r_{max}$]{
\includegraphics[width=2in]{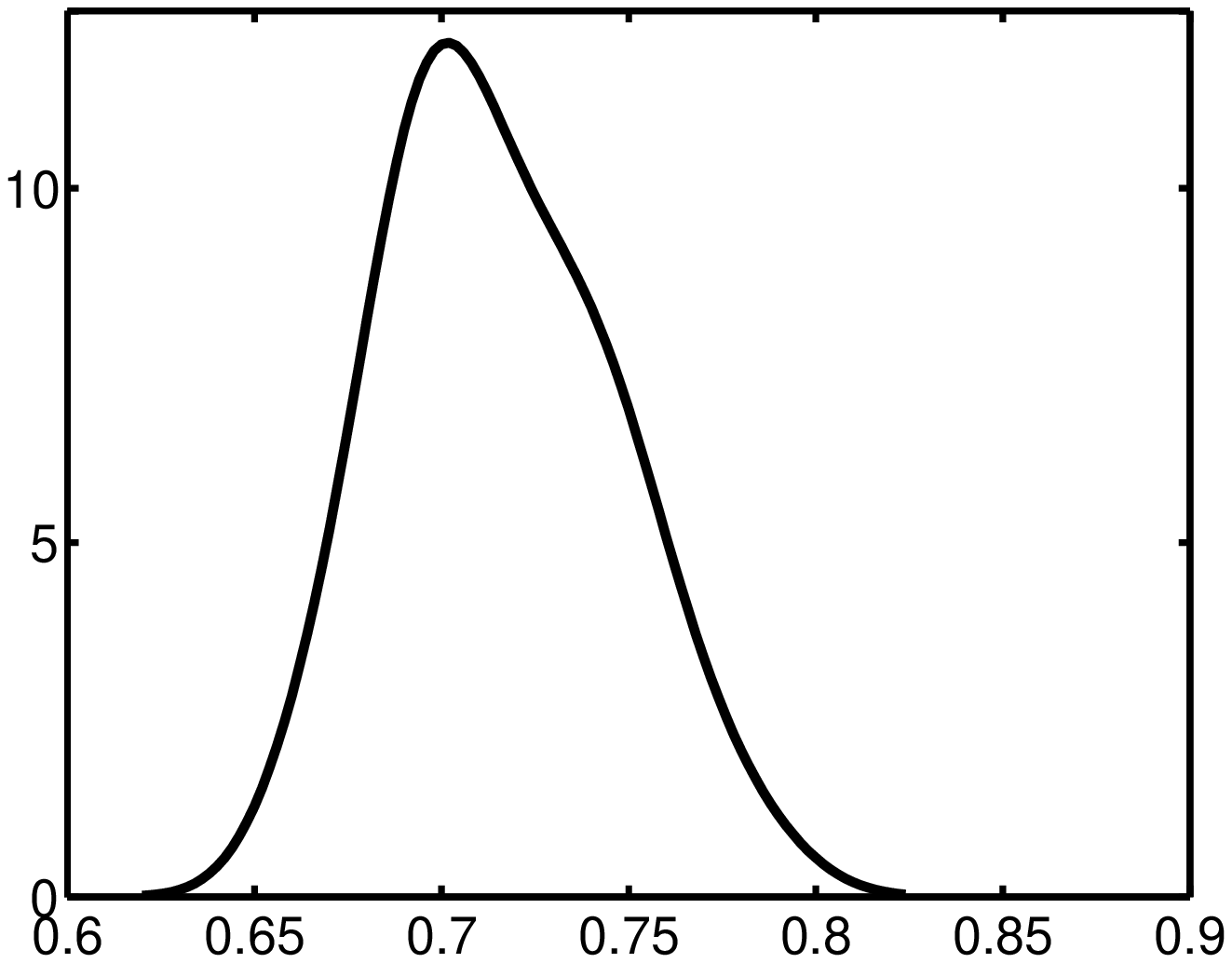}}
\caption{PDFs of $a,k_s$ and $r_{max}$ by Bayesian inference.}
\label{fig:infer_3d}
\end{figure}

Given the (approximated) response surface of $\eta$, $P$ and
$R_g$, we aim to infer the parameter $(a, k_s, r_{max})$ with the three
desirable target properties
%(\textcolor{green}{Xiu, please update the values below
%    using gPC term you've already computed})
\begin{equation}
\bm P^t=(\eta, R_g, P) = (4.457, 0.09862, 15.54).
\label{eq:poly3Dtarget}
\end{equation}
%-------------------------------------------------------------------------------
In order to infer $a, k_s, r_{max}$, we follow the framework in
\cite{Najm_UQ_2012_b} and employ the Bayesian theorem. In this
case, $\bm\theta=(a, k_s, r_{max})$ and we denote
$(G_1, G_2, G_3) = (\eta, R_g, P)$ as in \cite{Najm_UQ_2012_b}. The desirable
$\bm P^t$ can be written as
%-------------------------------------------------------------------------------
\begin{equation}
\label{eq:bayesian0}
\bm G_m = \{G_m^k\}_{k=1}^3, \qquad m=1,2,3,
\end{equation}
%-------------------------------------------------------------------------------
where $G_m^k$ represents the $k$-th replica for the $m$-th target property. A
direct Bayesian framework is employed to infer the posterior probability
density:
%-------------------------------------------------------------------------------
\begin{equation}
\label{eq:bayesian1}
\pi(\bm\theta | \{\bm G_m\}_{m=1}^3)\propto
\mathcal{L}(\{\bm G_m\}_{m=1}^3 | \bm\theta)q(\bm\theta),
\end{equation}
%-------------------------------------------------------------------------------
where $\mathcal{L}$ is the likelihood and $q$ is the prior. Due to the linear
relation between $\bm\theta$ and $\bx$ in Eq. \eqref{eq:para_uncertain2}, we can
equivalently consider the following posterior:
%-------------------------------------------------------------------------------
\begin{equation}
\label{eq:bayesian2}
\pi(\bx | \{\bm G_m\}_{m=1}^3)\propto
\mathcal{L}(\{\bm G_m\}_{m=1}^3 | \bx)q(\bx).
\end{equation}
%-------------------------------------------------------------------------------
We employ the same posterior as in \cite{Najm_UQ_2012_b}:
%-------------------------------------------------------------------------------
\begin{equation}
\label{eq:bayesian3}
\pi(\bx, \bm{\tilde\sigma}^2 | \{\bm G_m\}_{m=1}^3)\propto
\prod_{m=1}^3\prod_{k=1}^K
\dfrac{\exp\left(\dfrac{[G_m^k-\tilde X_m(\bx)]^2}{2\tilde\sigma_m^2}\right)}
{\sqrt{2\pi\tilde\sigma^2_m}} q(\tilde\sigma_m^2),
\end{equation}
%-------------------------------------------------------------------------------
where $\tilde X_m(\bx)$ is the value of the surrogate model at $\bx$ for the
$m$-th observable, {\color{black}i.e., we use the truncated polynomials to
    approximate observable $\tilde X_m$ (see Eq.~\eqref{eq:gpc2})} 
    and $\{\tilde\sigma_m^2\}_{m=1}^3$ are hyperparameters given a
Jeffreys prior of the form
%-------------------------------------------------------------------------------
\begin{equation}
q(\tilde\sigma_m^2) = \dfrac{1}{\tilde\sigma_m^2}, \qquad m=1,2,3.
\end{equation}
%-------------------------------------------------------------------------------
We use the Metropolis-Hasting Markov chain Monte Carlo (MCMC) method to sample
the posterior probability density Eq. \eqref{eq:bayesian3} by running $10^5$
steps and setting burn in period as $50,000$. The posterior distribution is
estimated by kernel smoothing function in MATALB. The results are presented in
Figure \ref{fig:infer_3d}. In this case, due to the property of
the model and the selection of the macroscopic observables, all three
PDFs are single modal. Thus, these parameters can be selected by the maximum
\textit{a posteriori} estimation probability (MAP) estimate based on the
inferred PDFs.
%-------------------------------------------------------------------------------
%-------------------------------------------------------------------------------
The set $\bm\theta$ we select is listed in Table \ref{tab:infer_par_3d}.
%-------------------------------------------------------------------------------
\begin{table}[!h]
\centering
\caption{Inferred parameters $\bm\theta$ for the 3D polymer melt
  system.}
\begin{tabular}{C{4em}*{3}{C{6em}}}
\hline \hline
& $a$ & $k_s$ & $r_{max}$ \\
\hline
$\bm\theta$ & 11.35 & 33.60 & 0.696 \\
\hline \hline
\end{tabular}
\label{tab:infer_par_3d}
\end{table}
%-------------------------------------------------------------------------------
The DPD simulation results $\bm P^{DPD}$ and the relative errors based on this
selected $\bm\theta$ are listed in Table \ref{tab:infer_3d}.
%-------------------------------------------------------------------------------
\begin{table}[!h]
\centering
\caption{Validation of the inferred parameters $\bm\theta$ for the 3D polymer
melt system. The DPD simulation results for three target properties are listed.
The relative error of each target property is also listed.}
\begin{tabular}{C{10em}*{3}{C{5em}}}
\hline \hline
 & $\eta$ & $R_g$ & $P$ \\
  \hline
$\bm P^{DPD}(\bm\theta)$ & 4.381 & 0.09893 & 15.40 \\
$|\bm P^t_i(\bm\theta)-\bm P^{DPD}_i|/|\bm P^t_i|$ &
1.7\% & 0.31\% & 0.89\%\\
\hline\hline
\end{tabular}
\label{tab:infer_3d}
\end{table}
%-------------------------------------------------------------------------------
It is clear that with the inferred parameters, the relative error for each
target property is $\mo(1\%)$ or smaller, which implies that our selection of
$\bm\theta$ is good. We also tested the above system with other sets of target
property values. By using the inferred parameters, the present mesoscopic model yields consistent results with the prescribed target property values.

\begin{rem}
There are different approaches to infer the model parameters and here we employ a
simple and direct one in our demonstration since we focus on how to construct
the surrogate model fast. More sophisticated methods, e.g., adaptive
MCMC \cite{HaarioST01}, DRAM \cite{HaarioLMS06}, TMCMC \cite{Ching_Chen_2007,Pan_Kou_JCP_2012}
can be applied once the surrogate models are constructed and the desirable
properties are given.
\end{rem}

%-------------------------------------------------------------------------------

\subsubsection{6D polymer model}
\label{sec:6D_model}

The setup of the 6D polymer model is summarized as follows:
\begin{itemize}  
\item {\em Mesoscopic system}: polymer melt model defined by Eq. (\ref{eq:DPD_eq2}) 
and Eq. (\ref{eq:fene}) with
number density $n = 3.0$, $N_b = 5$ and $k_BT = 1.0$.
\item {\em Target properties for parameter inference}: viscosity at shear-rate
$0.06, 0.07, 0.08$ ($\eta_{0.06},\eta_{0.07},\eta_{0.08}$), diffusivity ($D$),
average of radius of gyration ($R_g$) and pressure ($P$) with values specified
by Eq. (\ref{eq:polymertarget}).
\item {\em Inferred model parameters}: $a(\xi_1)$, 
$\gamma(\xi_2)$, $k(\xi_3)$, $r_c(\xi_4)$, 
$k_s(\xi_5)$ and $r_{max}(\xi_6)$ with parameter confidence range
specified by Eq. (\ref{eq:para_uncertain3}).
\end{itemize}

We study the full polymer melt system with 6 parameters discussed in
Section \ref{subsec:poly_melt} with $N_b = 5$, $n = 3.0$ and $k_B T = 1.0$,
given the parameter space $(a,\gamma,k,r_c,k_s,r_{max})$ defined by
\begin{equation}
\begin{aligned}
 a(\xi_1) &= 25.0 + \sigma_{a}\xi_1, & \gamma(\xi_2) &= 8.0 + \sigma_{\gamma}\xi_2,\\
 k(\xi_3) &= 0.25 + \sigma_{k}\xi_3, & r_c(\xi_4) &= 1.35  + \sigma_{r_c}\xi_4,\\
 k_s(\xi_5) &= 50.0 + \sigma_{k_s}\xi_5, & r_{max}(\xi_6) &= 0.85 + \sigma_{r_{max}}\xi_6,
\end{aligned}
\label{eq:para_uncertain3}
\end{equation}
where
$(\sigma_{a}, \sigma_{\gamma}, \sigma_{k}, \sigma_{r_c}, \sigma_{k_s}, \sigma_{r_{max}}) = (15.0, 4.0, 0.1, 0.05, 30.0, 0.25)$
and $\bx = (\xi_1, \xi_2, ...,\xi_6)$ are \textit{i.i.d} uniform random variables
on $[-1,1]$. We aim to infer all 6 parameters in the DPD model given
a sufficient number of macroscopic observable
properties. More precisely, we set
$\bm\theta=(a,\gamma,k,r_c,k_s,r_{max})$,
and target  the following 6 properties: viscosity at shear-rate
$0.06, 0.07, 0.08$ ($\eta_{0.06},\eta_{0.07},\eta_{0.08}$), diffusivity ($D$),
average of radius of gyration ($R_g$) and pressure ($P$). 

\begin{figure}[!h]
\centering
\subfigure[$a$]{
\includegraphics[width=2.5in]{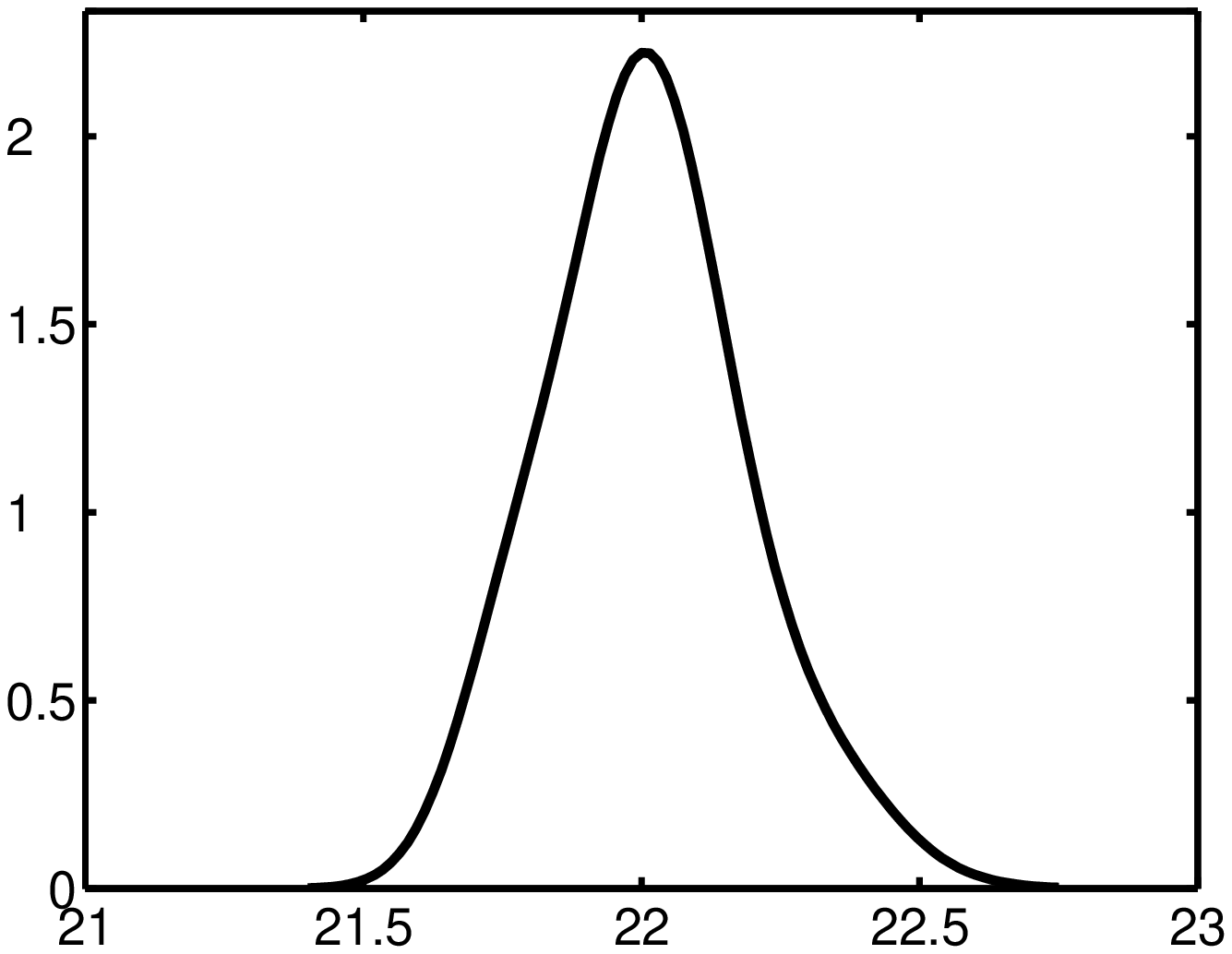}} \quad
\subfigure[$r_c$]{
\includegraphics[width=2.5in]{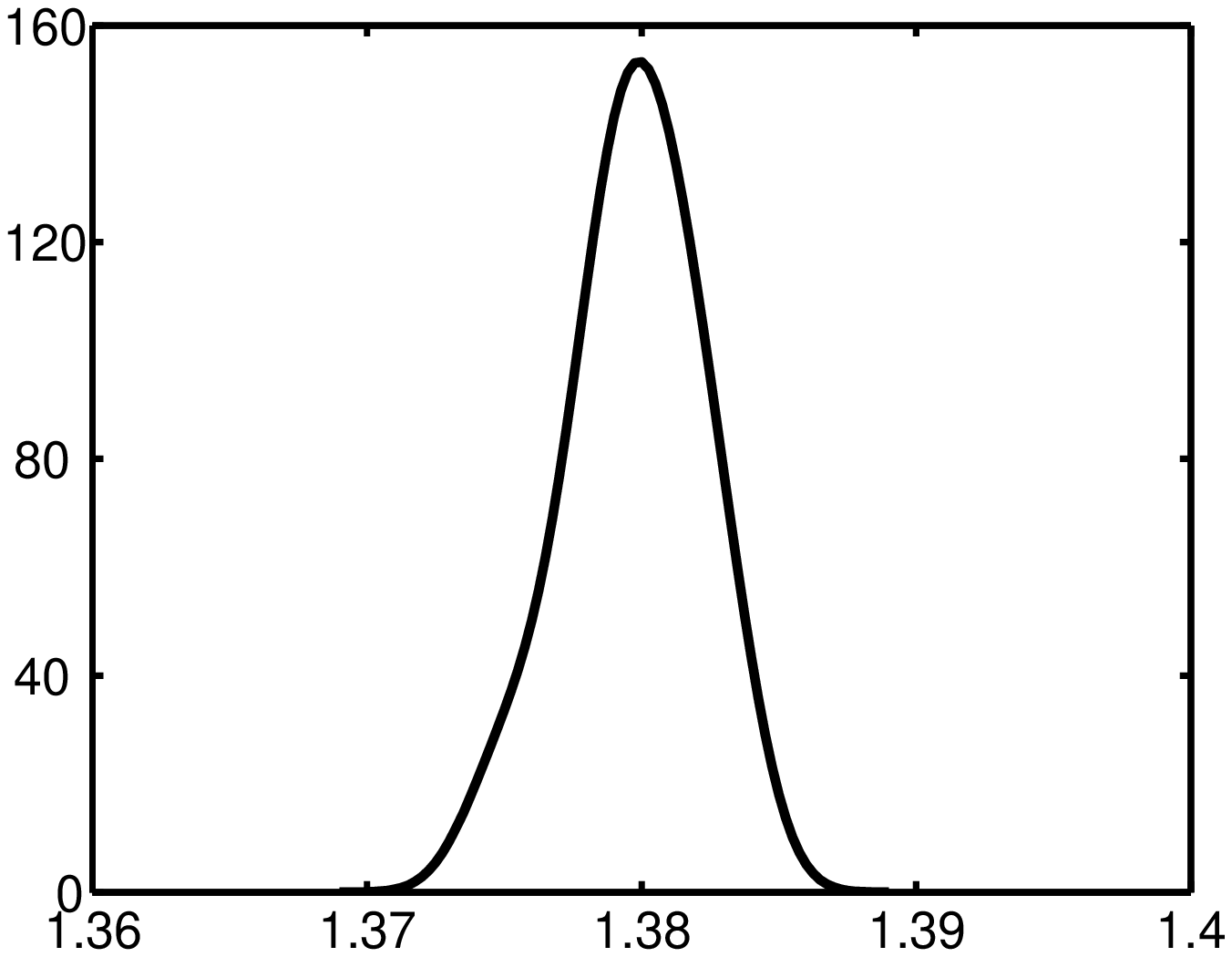}} \\
\subfigure[$k_s$]{
\includegraphics[width=2.5in]{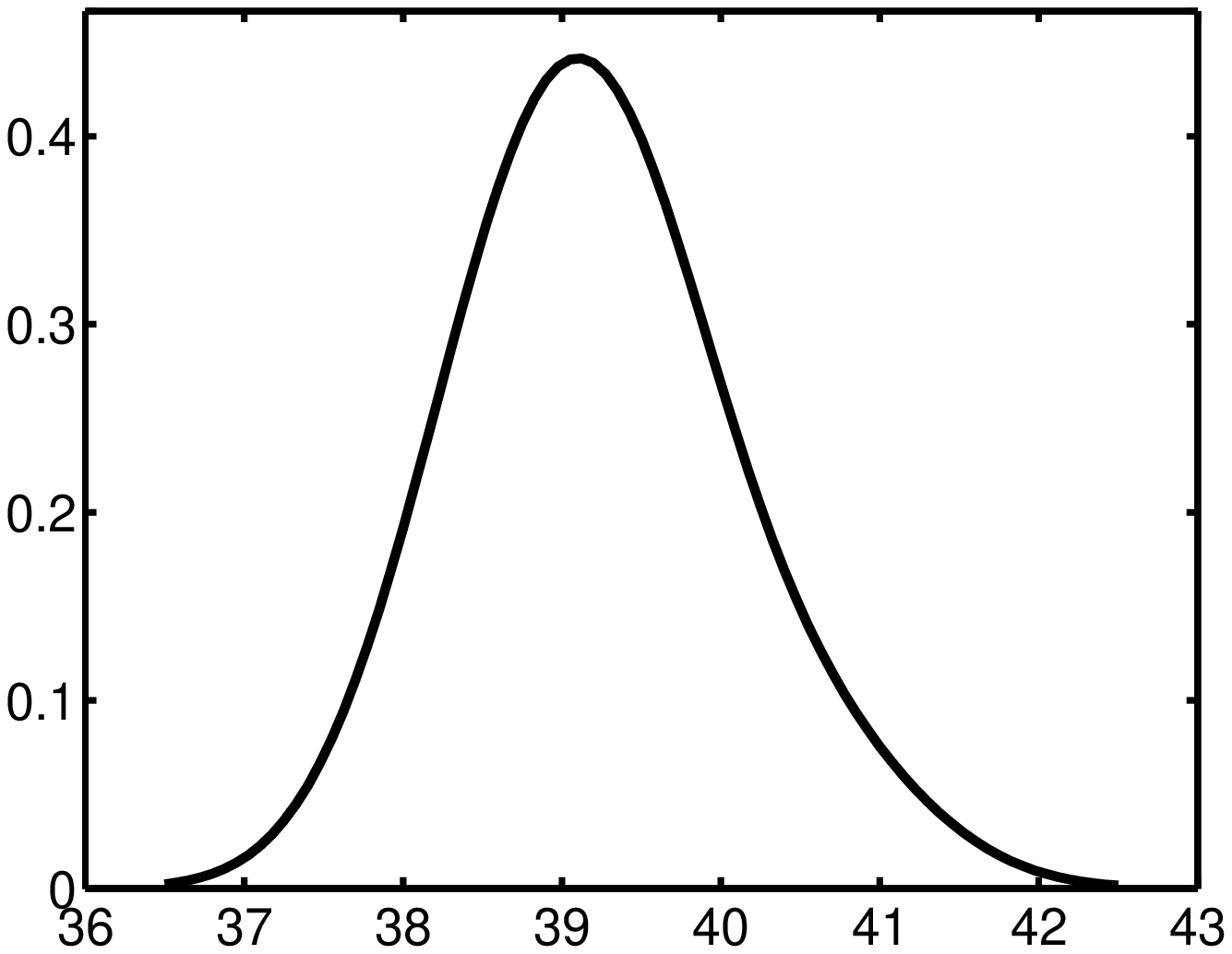}} \quad
\subfigure[$r_{max}$]{
\includegraphics[width=2.5in]{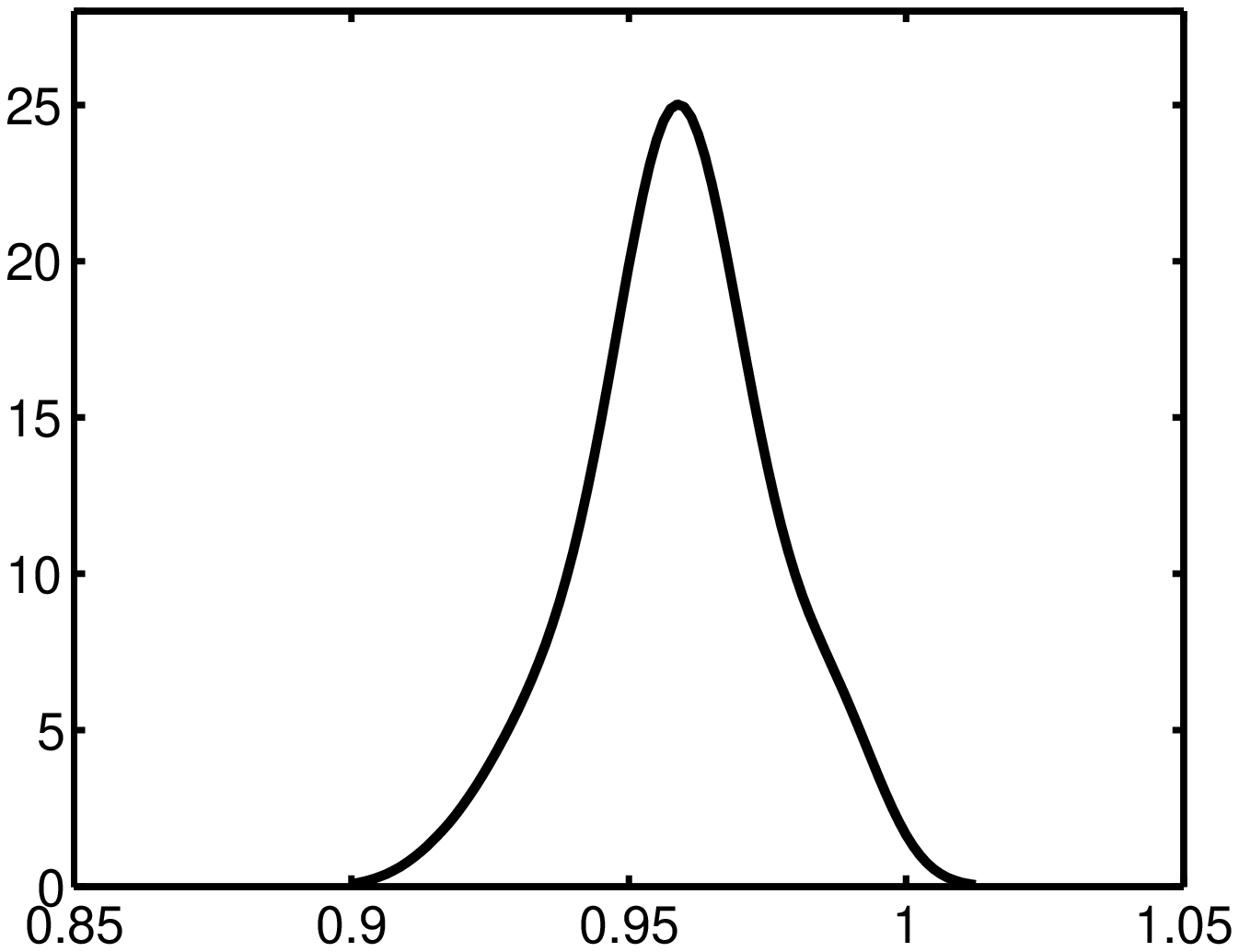}}
\caption{PDFs of $a,r_c,k_s$ and $r_{max}$ by Bayesian inference.}
\label{fig:infer6d_1}
\end{figure}

\begin{figure}[!h]
\centering
\includegraphics[width=3.5in]{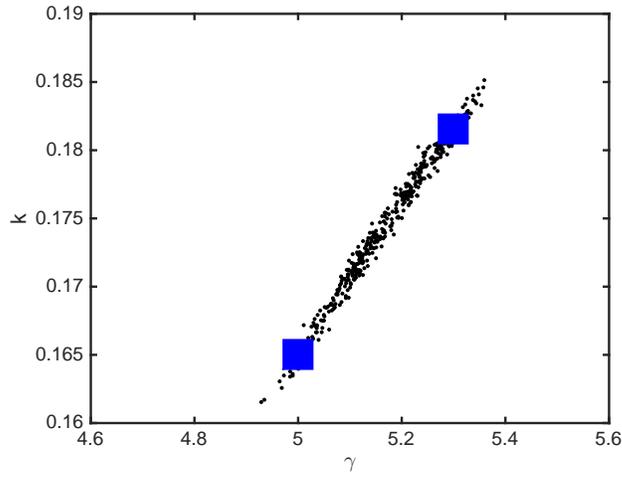}
\caption{MCMC samples of $(\gamma, k)$. The two square points are specified in
Table \ref{tab:infer_par_6d}.}
\label{fig:infer6d_2}
\end{figure}

In this test we construct a surrogate model with up to third-order Legendre polynomials as well
as $46$ fourth-order Legendre polynomials based on $110$ samples of DPD
simulations. We denote
$\bm P^t=(G_1,G_2,G_3,G_4,G_5,G_6)=(\eta_{0.06},\eta_{0.07},\eta_{0.08},D,R_g,P)$,
and similar to Equation \eqref{eq:bayesian0} the replicas of the properties are
written as
%-------------------------------------------------------------------------------
\begin{equation}
\label{eq:bayesian4}
\bm G_m = \{G_m^k\}_{k=1}^3, \qquad m=1,2,3,4,5,6.
\end{equation}
%-------------------------------------------------------------------------------
Then we employ Eq. \eqref{eq:bayesian3} again by changing the upper
range of $m$ to $6$ since we have 6 properties. We set the desirable target
property as
\begin{equation}
\label{eq:polymertarget}
\bm P^t = (29.12, 27.96, 26.88, 0.003563, 0.1726, 73.69).
\end{equation}
Figure \ref{fig:infer6d_1}
presents the inference results for $\xi_1,\xi_4,\xi_5,\xi_6$ (i.e., $a, r_c, k_s, r_{max}$). We can see that
these $4$ parameters can be inferred by MAP. However, for $\xi_2$ and $\xi_3$ (i.e., $\gamma$  and $k$)
we obtain bi-modal PDFs. Hence, we plot MCMC samples of $(\gamma, k)$ in
Figure \ref{fig:infer6d_2} to investigate the correlation between these two
parameters. This plot reveals a strong correlation between $\gamma$ and $k$. We
also plot similar graphs to
investigate possible correlations between
other pairs of parameters (not presented here) but we do not
observe such correlations.

%-------------------------------------------------------------------------------
%-------------------------------------------------------------------------------
%-------------------------------------------------------------------------------
Therefore, in order to verify the inference results, we select two parameter
sets for test (see Table \ref{tab:infer_par_6d}). The locations of $(\gamma, k)$
are presented in Figure \ref{fig:infer6d_2} with square points.
\begin{table}[!h]
\centering
\caption{Two sets of inferred parameters $\bm\theta$ for the 6D polymer melt
  system.}
\begin{tabular}{C{6em}*{6}{C{4em}}}
\hline \hline
& $a$ & $\gamma$ & $k$ & $r_c$ & $k_s$ & $r_{max}$ \\
\hline
$\bm\theta^1$ & 22 & 5.00 & 0.165 & 1.38 & 39.2 & 0.9575 \\
$\bm\theta^2$ & 22 & 5.30 & 0.182 & 1.38 & 39.2 & 0.9575 \\
\hline \hline
\end{tabular}
\label{tab:infer_par_6d}
\end{table}
%-------------------------------------------------------------------------------
Notice that here we fixed $a,r_c, k_s, r_{max}$ and select two different
sets of $(\gamma, k)$ according to Figure \ref{fig:infer6d_2}. We obtain two
sets of target properties from the DPD simulation as presented in
Table \ref{tab:infer_6d}.  The relative error of each target property is also
listed. We observe that the relative error is $\mo(1\%)$ or smaller, hence, the
selections of both parameter sets are good. This implies that in our DPD model
we only need to keep $\gamma$ or $k$ and use the correlation revealed in
Figure \ref{fig:infer6d_2} to set the other parameter. Thus, we achieve a
\emph{model reduction} by decreasing the degree of freedom of the model by one. In
other words, the Bayesian inference result implies that, in this polymer melt
system, we only need five parameters in our DPD model to capture the $6$ target
properties we need.
%-------------------------------------------------------------------------------
\begin{table}[!h]
\centering
\caption{Validation of the inferred parameters $\bm\theta^1$ and $\bm\theta^2$
for the 6D polymer melt system. The DPD simulation results for $6$ target
properties with different $\bm\theta$ are listed. The relative error of each
target property is also listed.}
\begin{tabular}{C{10em}*{6}{C{4em}}}
\hline \hline
  & $\eta_{0.06}$ & $\eta_{0.07}$ & $\eta_{0.08}$ & $D$ & $R_g$ & $P$ \\
  \hline
$\bm P^{DPD}(\bm\theta^1)$ & 29.38 & 28.05 & 26.83 & 0.00358 & 0.1722 & 73.65\\
$|\bm P^t_i(\bm\theta^1)-\bm P^{DPD}_i|/|\bm P^t_i|$ &
0.89\% & 0.32\% & 0.19\% & 0.48\% & 0.23\% & 0.05\%\\
\hline
$\bm P^{DPD}(\bm\theta^2)$ & 29.47 & 28.13 & 26.90 & 0.00354 & 0.1722 & 73.65\\
$|\bm P^t_i(\bm\theta^2)-\bm P^{DPD}_i|/|\bm P^t_i|$ &
1.20\% & 0.61\% & 0.07\% & 0.65\% & 0.23\% & 0.05\%  \\
\hline\hline
\end{tabular}
\label{tab:infer_6d}
\end{table}
%-------------------------------------------------------------------------------
\begin{rem}
In Table \ref{tab:infer_6d}, we notice that the relative errors of
different bulk properties varies. For example, the error of the pressure is
smaller than other bulk properties. This is because: (1) the thermal noise has
very little impact on the pressure; (2) the surrogate model for the pressure is
more accurate than other properties due to the sparsity of its gPC coefficients;
(3) the pressure is less sensitive with respect to the parameters within the
range of inferred parameter values.
\end{rem}

\begin{figure}[!h]
\center
\subfigure[Viscosity]{
\includegraphics[width=3in]{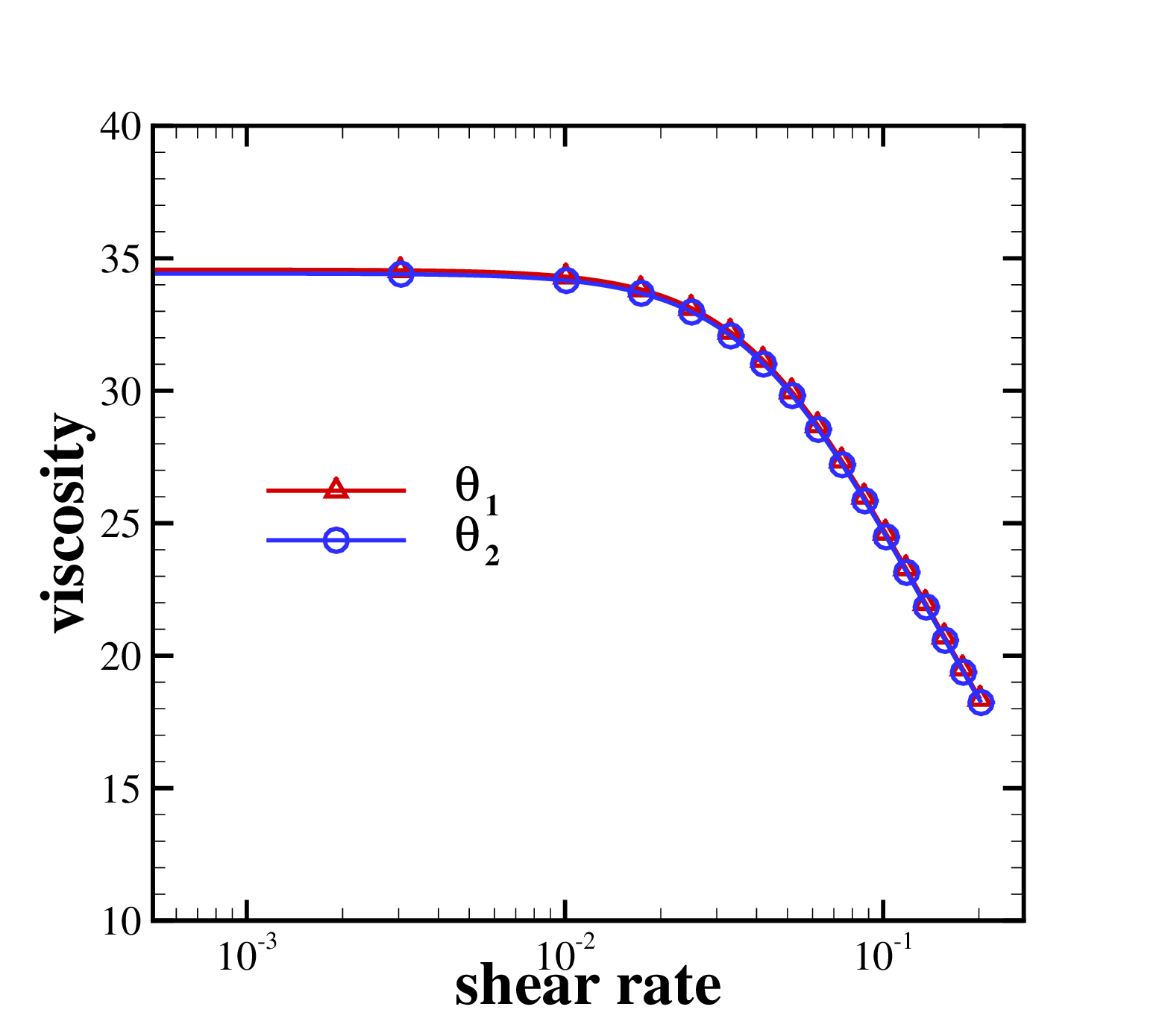}}
\subfigure[MSD]{
\includegraphics[width=3in]{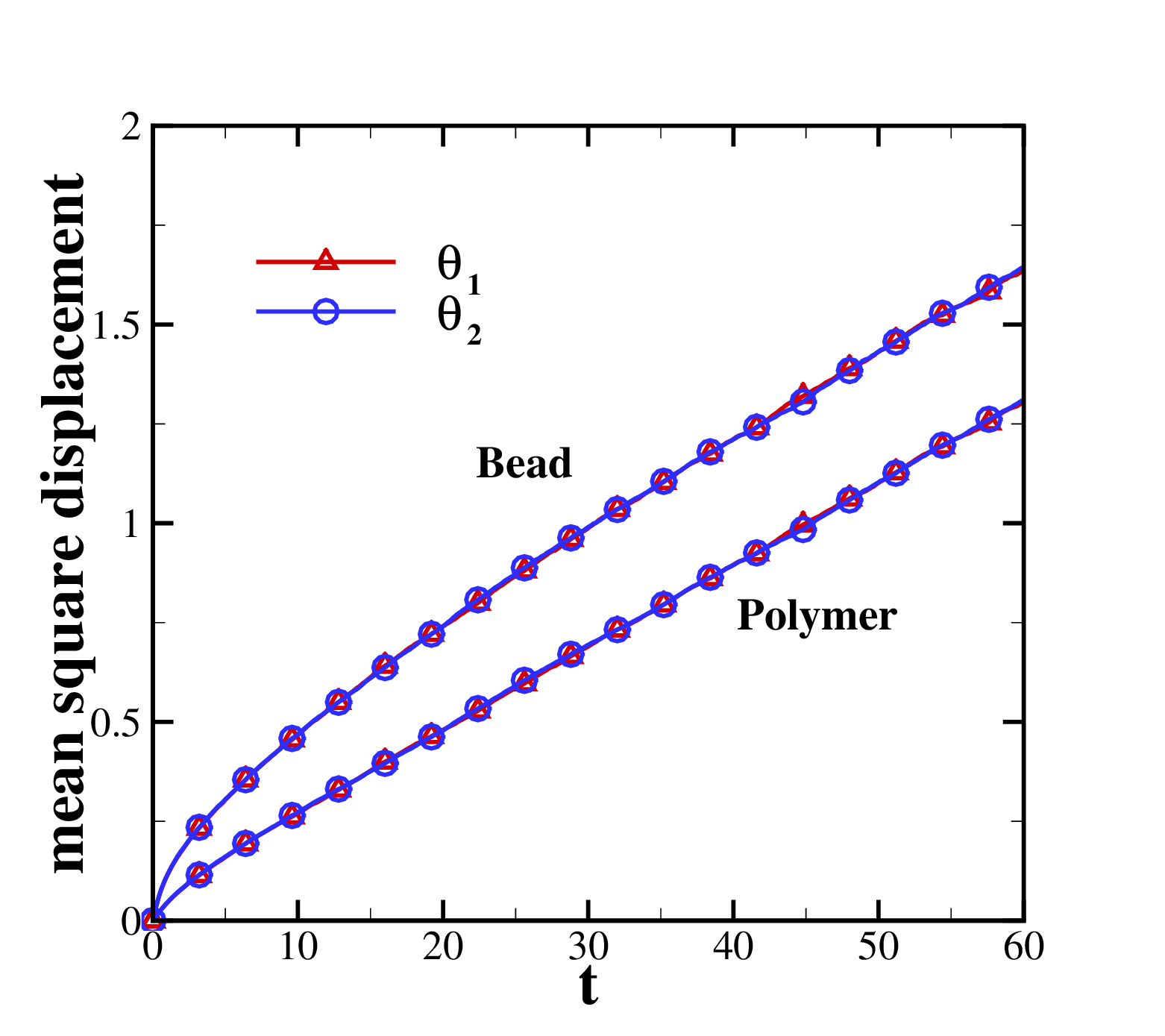}} \\
\subfigure[Bulk relaxation time]{
\includegraphics[width=3in]{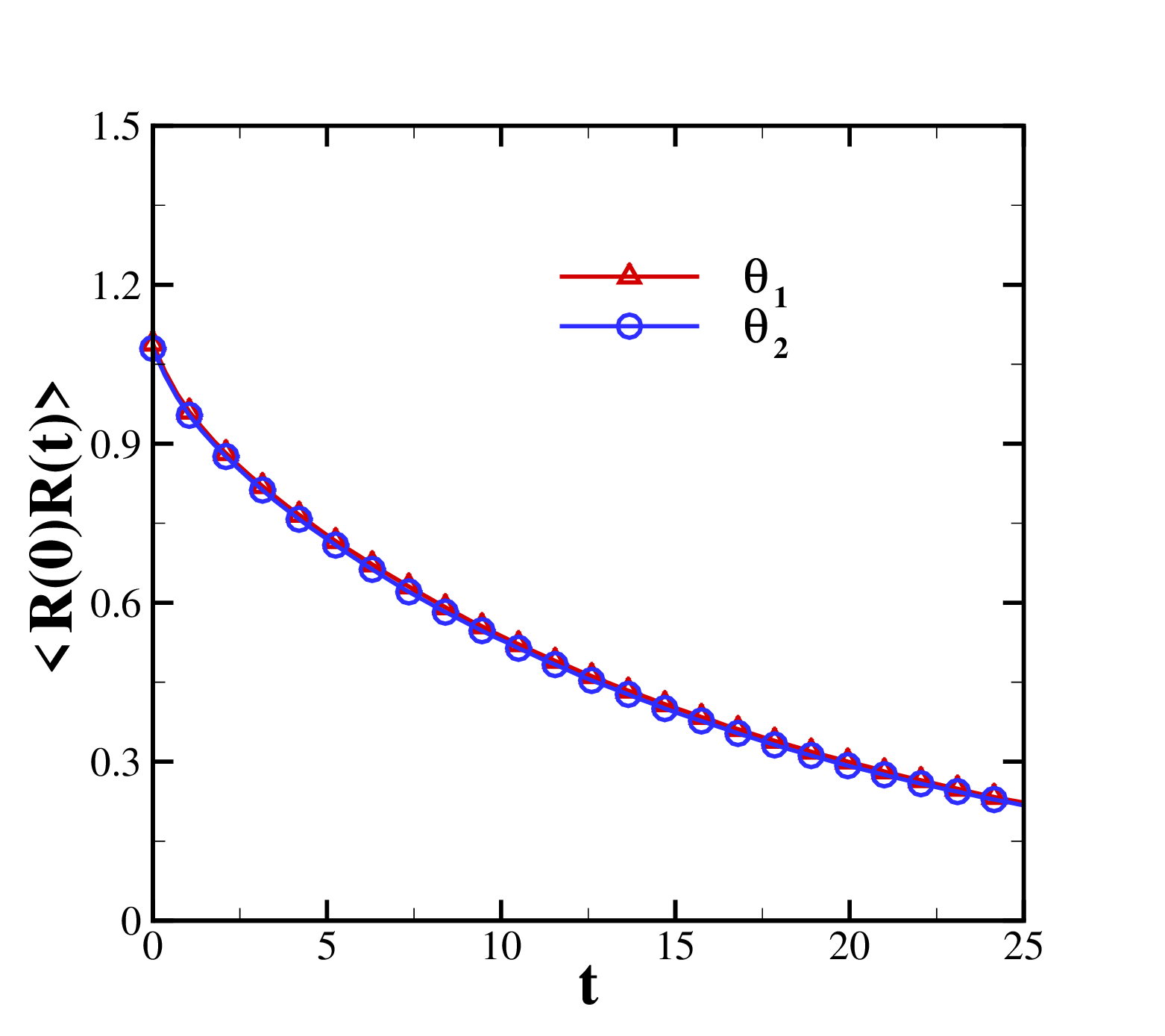}}
\subfigure[Poiseuille flow relaxation time]{
\includegraphics[width=3in]{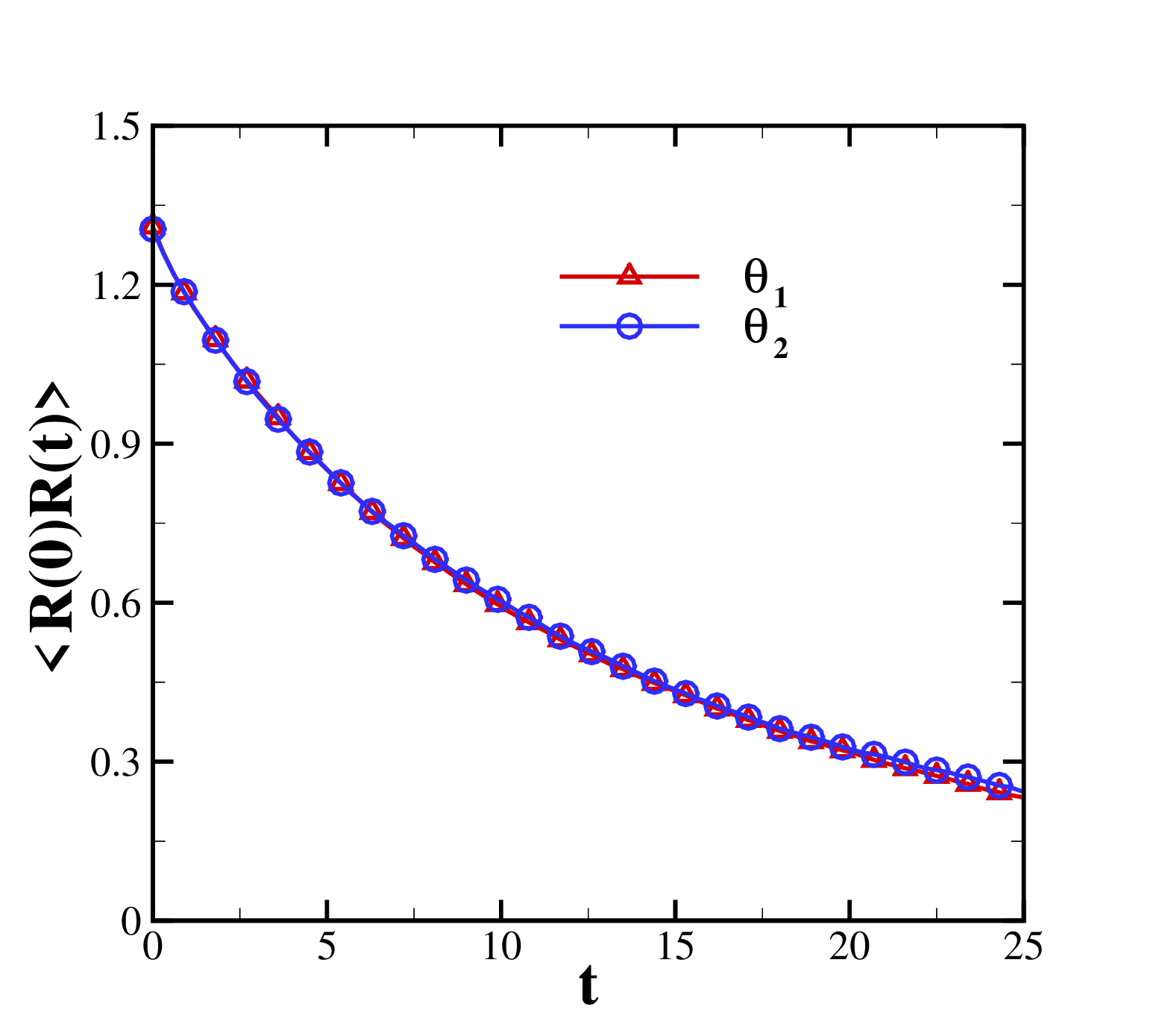}}
\caption{The shear viscosity, mean square displacement both individual DPD bead
and center of mass of polymer (MSD), bulk relaxation time and relaxation time
from reverse Poiseuille flow for the polymer melt systems with the parameter set
$\bm\theta^1$ and $\bm\theta^2$ in Table \ref{tab:infer_par_6d}.}
\label{fig:degeneracy}
\end{figure}

The Bayesian inference results
indicate that there exists certain parametric redundancy
{\em we were not aware of} when constructing the mesoscopic model for the polymer melt
system in Section \ref{sec:dpd}. In order to validate this hypothesis, we 
further investigate the other dynamic
properties of the polymer melt system with two different parameter sets
$\bm\theta^1$ and $\bm\theta^2$ in Table \ref{tab:infer_par_6d}.
The various dynamic
properties of the polymer melt system are shown in Figure \ref{fig:degeneracy}.
First, we compute the full shear rate dependent viscosity following the
method explained in Section \ref{sec:vis_rheometer}. The two response curves
show good agreement within the whole shear rate regime. Next, we consider the
bulk diffusivity by computing the mean square displacement of the individual
DPD bead as well as the center of mass of individual polymer. Again the two
parameter sets generate consistent results. Moreover, we consider the relaxation
time of individual polymer determined by the time correlation of the end-to-end
vector of individual polymers, e.g., $\left<\mathbf{R}(0)\mathbf{R}(t)\right>$,
where $\mathbf{R}(t)$ represent the instantaneous end-to-end vector of an
individual polymer under equilibrium state. The simulation results agree well
with each other. Finally, we consider the relaxation process of the polymer melt
system {computed independently} from a periodic opposite pressure driven flow as sketched in
Figure \ref{fig:simulation_box}. The initial state is obtained by applying equal but
opposite body force $g = 0.2$ on individual DPD particles. At $t = 0$, we remove
the body force and compute the evolution of individual polymer using
$\left<\mathbf{R}(0)\mathbf{R}(t)\right>$. Figure \ref{fig:degeneracy} validates
that the two parameter sets result in the same simulation results.
%-------------------------------------------------------------------------------
%-------------------------------------------------------------------------------
We have also conducted the above study for other parameter sets, generating
similar consistent results. {Taken together, all} these results demonstrate that there exists
parameter degeneracy in the mesoscopic model of the polymer melt system, which
may not be easily identified in a straightforward way.
Therefore, the present model can be further simplified {by eliminating 
parameter redundancies} according to the correlation function identified from the above analysis.

\begin{figure}
\center
\includegraphics[width=5in]{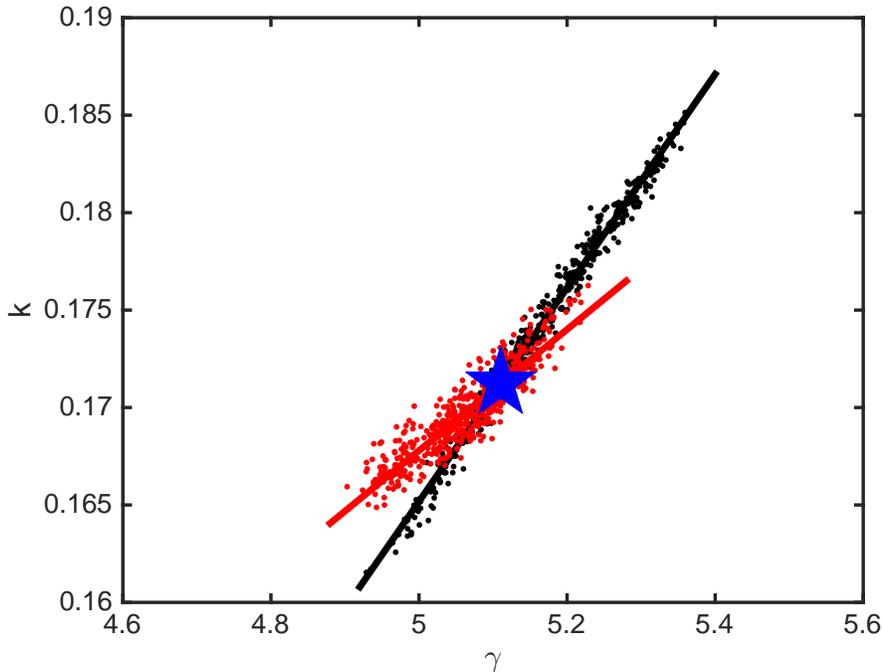}
\caption{{Parameters $\gamma$ and $k$ inferred from simulation results
 of polymer melts with density $3.0$ (black dots) and $5.0$ (red dots), respectively. 
 When multiple density regimes are considered, parameter degeneracy is eliminated.} 
 The location of the  {star symbol} is $\gamma=5.11, k=0.171$.}
\label{fig:parameter_decouple}
\end{figure}

{ Moreover, we emphasize that the parameter degeneracy identified in the 
present system further depends on target properties and parameter confidence range
we aim to recover. For instance, if we assume that the mesoscopic force field of the
polymer melt system weakly depends on the polymer number density (e.g., 
many-body effect is weak, see \cite{Lei_Cas_2010} for details discussion) and infer the
model parameters by targeting properties $\bm P^t$ following Eq. (\ref{eq:polymertarget}) 
for number density $n = 3$ and $n = 5$ simultaneously, we are able to infer parameter
$(\xi_2, \xi_3)$ with unique set of values, as shown in Fig. \ref{fig:parameter_decouple}.
However, this result is not
unexpected since the spirit of coarse-graining is to utilize a \textit{simple}
formulation to recover \textit{less} number of target properties. The more
properties we target, the more parameters we need to incorporate into the 
mesoscopic model. In practice, we need to calibrate the model parameter and the formulation
within specific parameter confidence range.
}

\begin{rem}
Based on the framework presented in this paper, we may obtain several sets of
parameters that are able to capture the target properties. This is not only
because there may be correlations between the parameters as studied in the 6D
polymer melt system, but it may also be because there are indeed several sets of
suitable parameters with no correlation between them given a
wide parameter confidence range. Therefore, it is possible
with a different approach to obtain the posterior in the Bayesian inference,
we may obtain different sets of parameters. Hence, we can introduce more
target properties to select the optimal parameter, or tune 
the parameter confidence range according to empirical values.
\end{rem}

%===============================================================================

\subsubsection{eDPD model}
\label{sec:eDPD}

{The setup of the eDPD model is summarized as below:}
\begin{itemize}  
\item {{\em Mesoscopic system}: non-isothermal liquid water system 
defined by Eq. (\ref{eq:eDPD1}), Eq. (\ref{equ:heat1}), Eq. (\ref{equ:heat2}) 
and Eq. (\ref{equ:heat3}) with reduced temperature $T = 0.91$ to $1.2433$, $n=4.0$, $a_{ij}=75k_BT/n$, 
$\gamma=4.5$, $r_c=1.58$}.
\item {{\em Target properties for parameter inference}: temperature
dependent viscosity $\eta_{0.91}, \eta_{1.1}, \eta_{1.2433}$
and viscosity $D_{0.91}, D_{1.1}, D_{1.2433}$ with values specified
by Eq. (\ref{eq:edpdtarget}).}
\item {{\em Inferred model parameters}: Coefficients $C_1, C_2,
C_3, C_4$ with parameter confidence range
specified by Eq. (\ref{eq:para_edpd}).}
\end{itemize}

\begin{figure}[!h]
\centering
\subfigure[$C_1$]{
\includegraphics[width=2.5in]{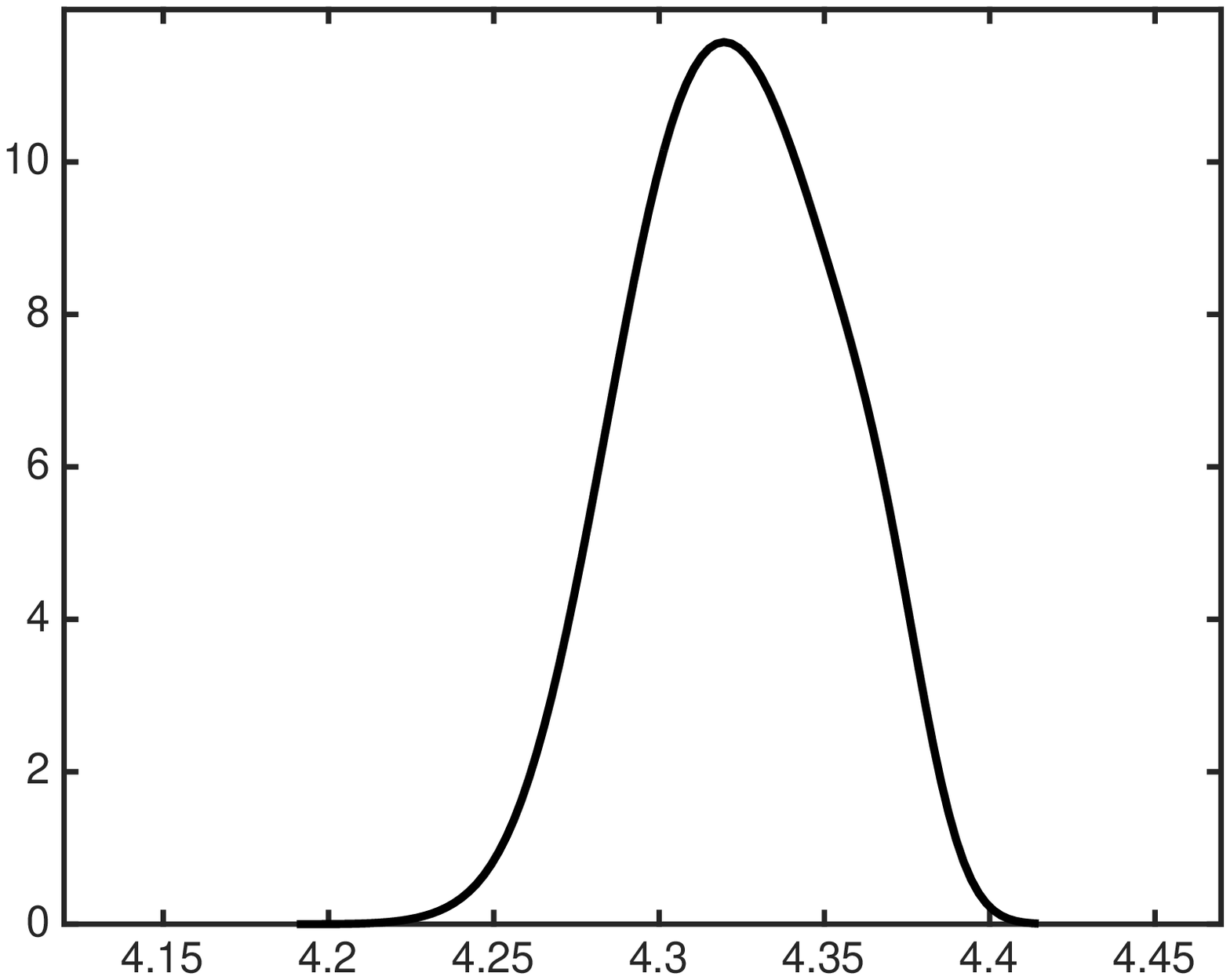}} \quad
\subfigure[$C_2$]{
\includegraphics[width=2.5in]{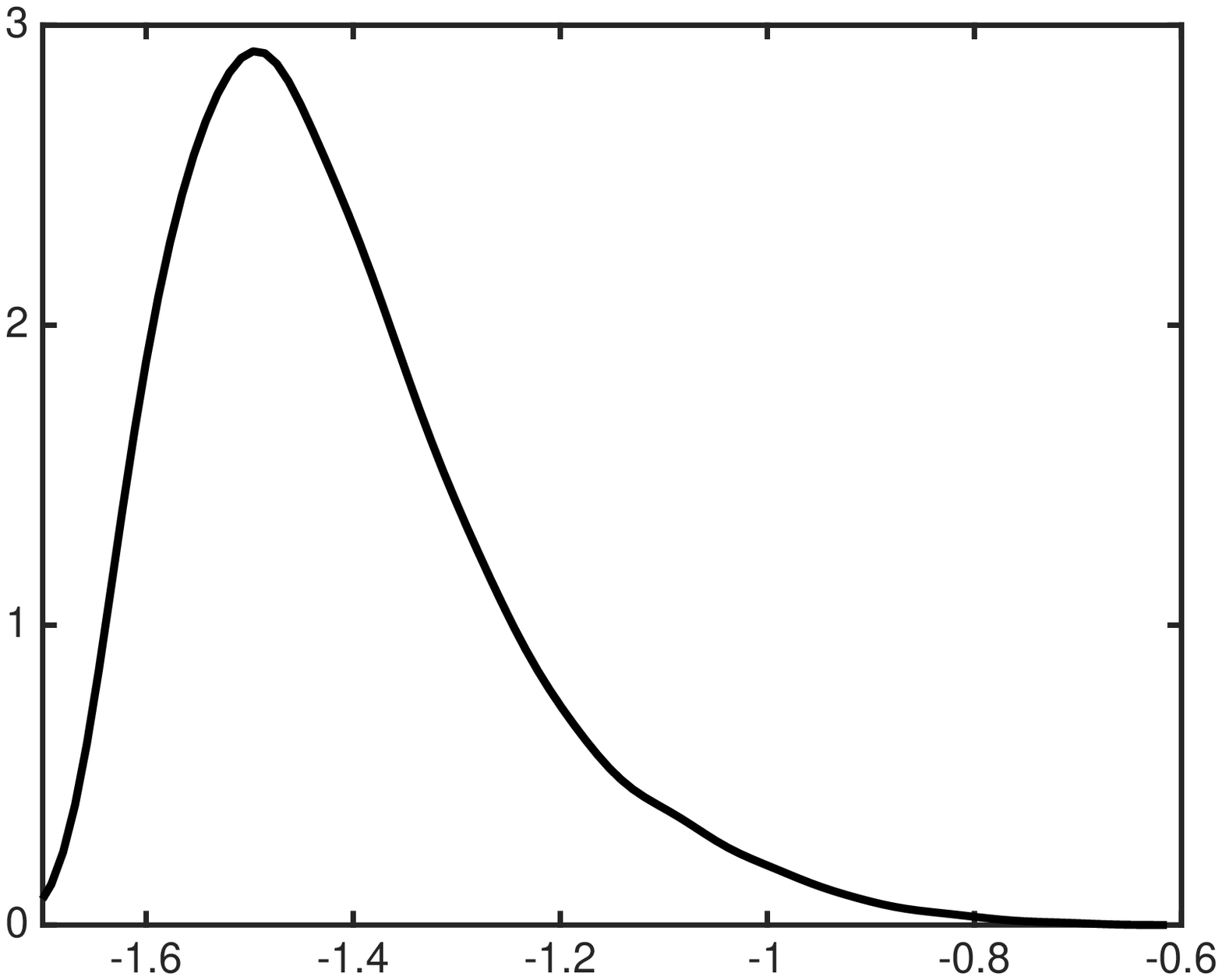}} \\
\subfigure[$C_3$]{
\includegraphics[width=2.5in]{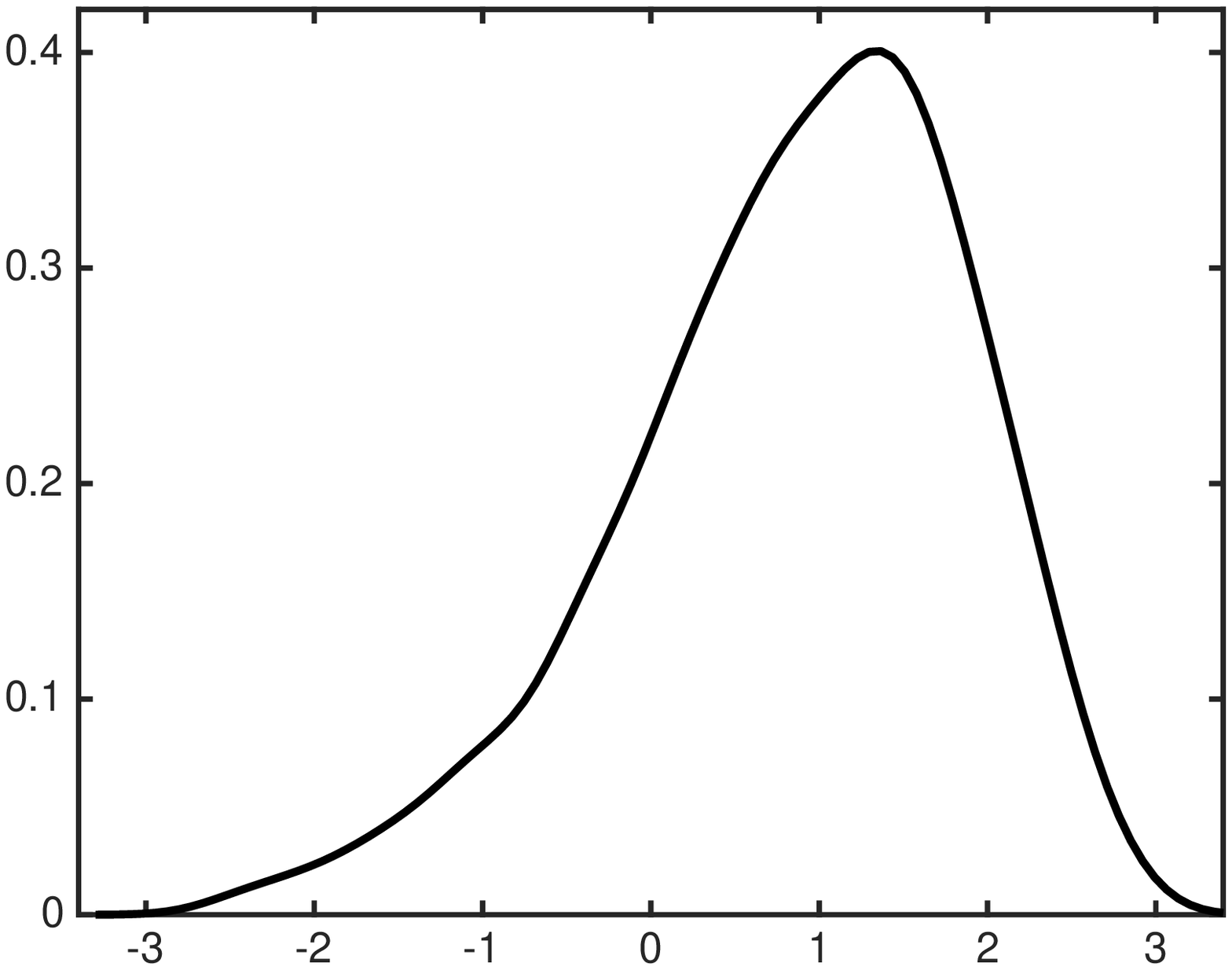}} \quad
\subfigure[$C_4$]{
\includegraphics[width=2.5in]{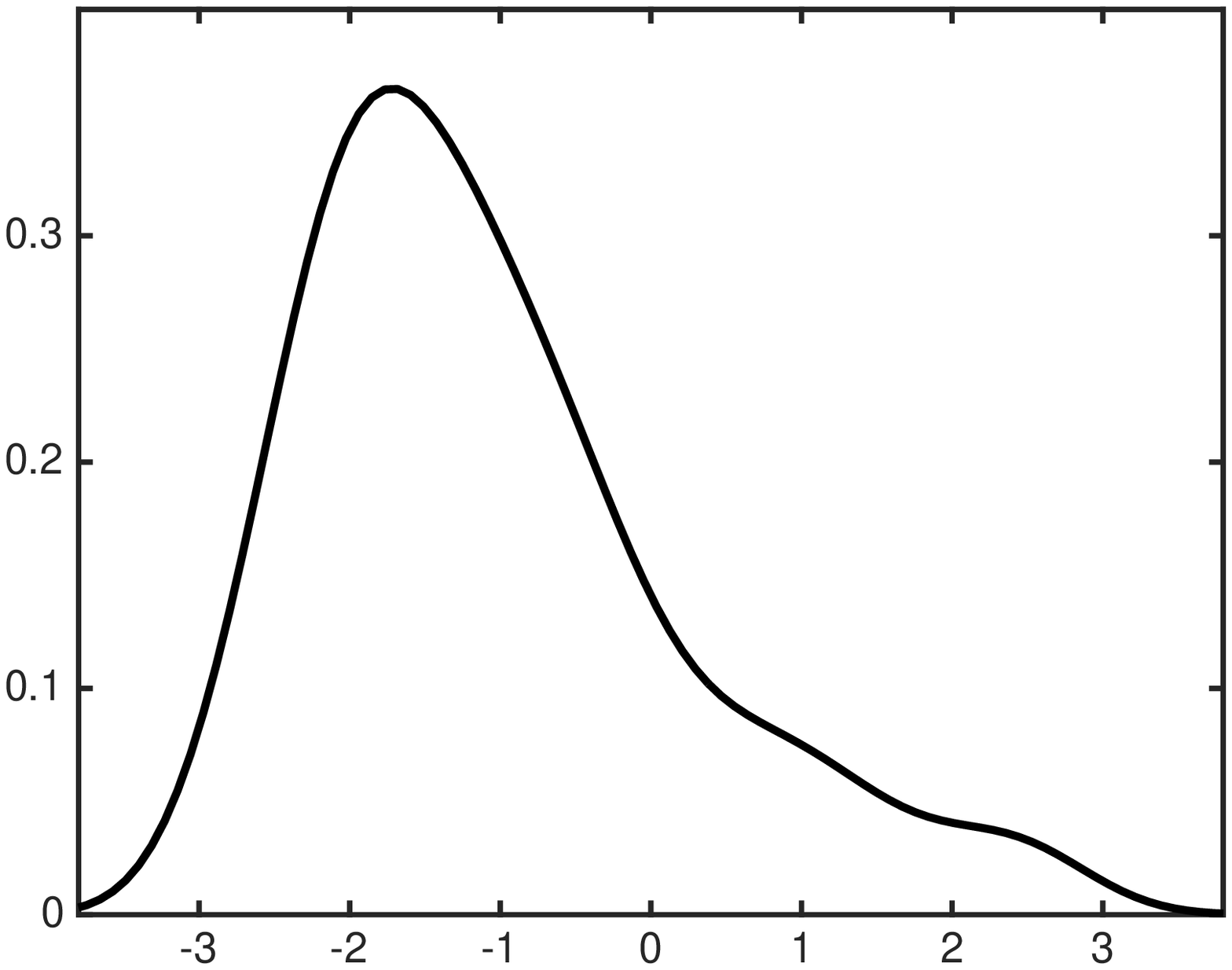}}
\caption{PDFs of $C_i, i=1,2,3,4$ by Bayesian inference.}
\label{fig:eDPD_infer}
\end{figure}

\begin{figure}[!h]
\subfigure[Diffusivity]{
\includegraphics[width=3in]{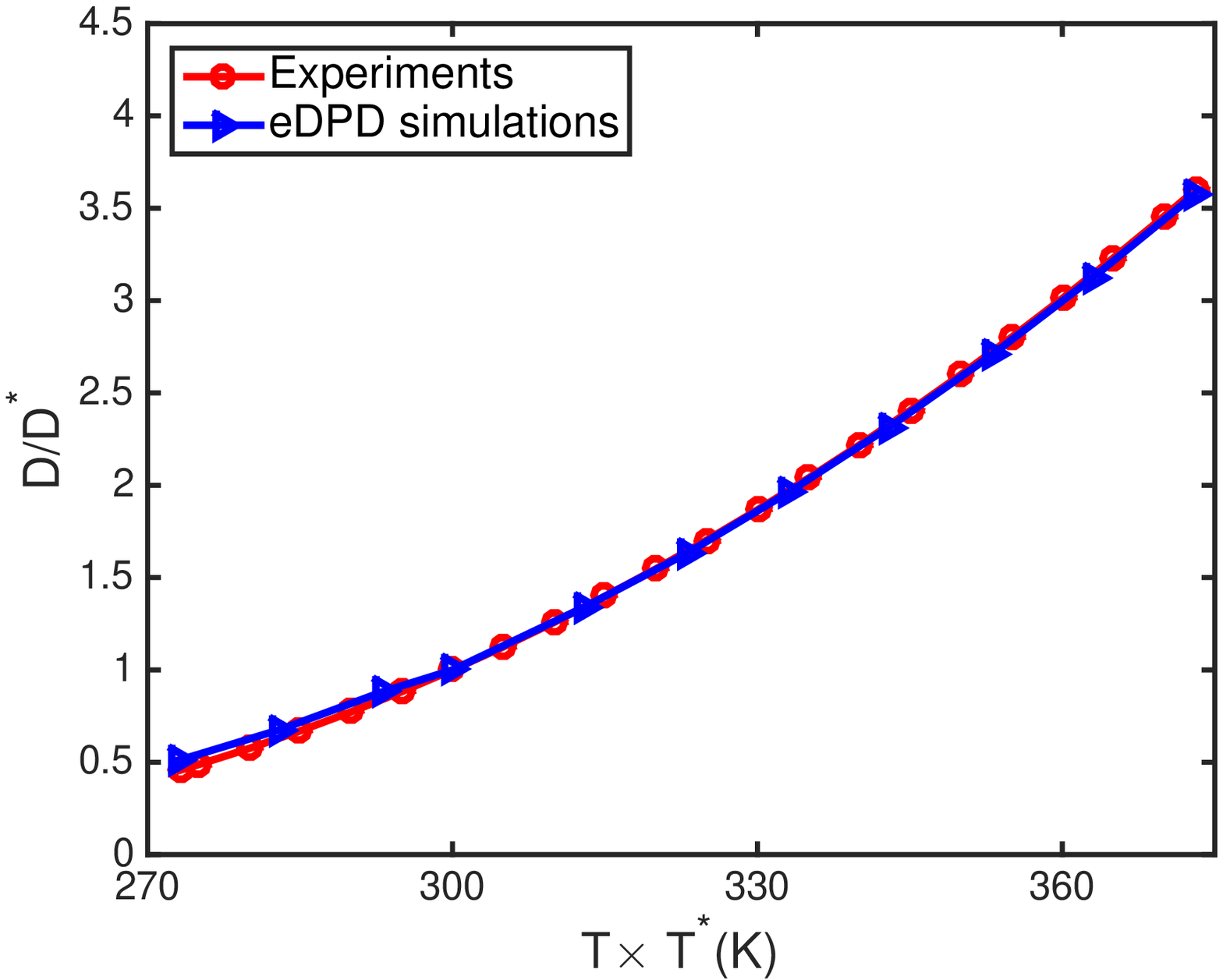}} \quad
\subfigure[Viscsity]{
\includegraphics[width=3in]{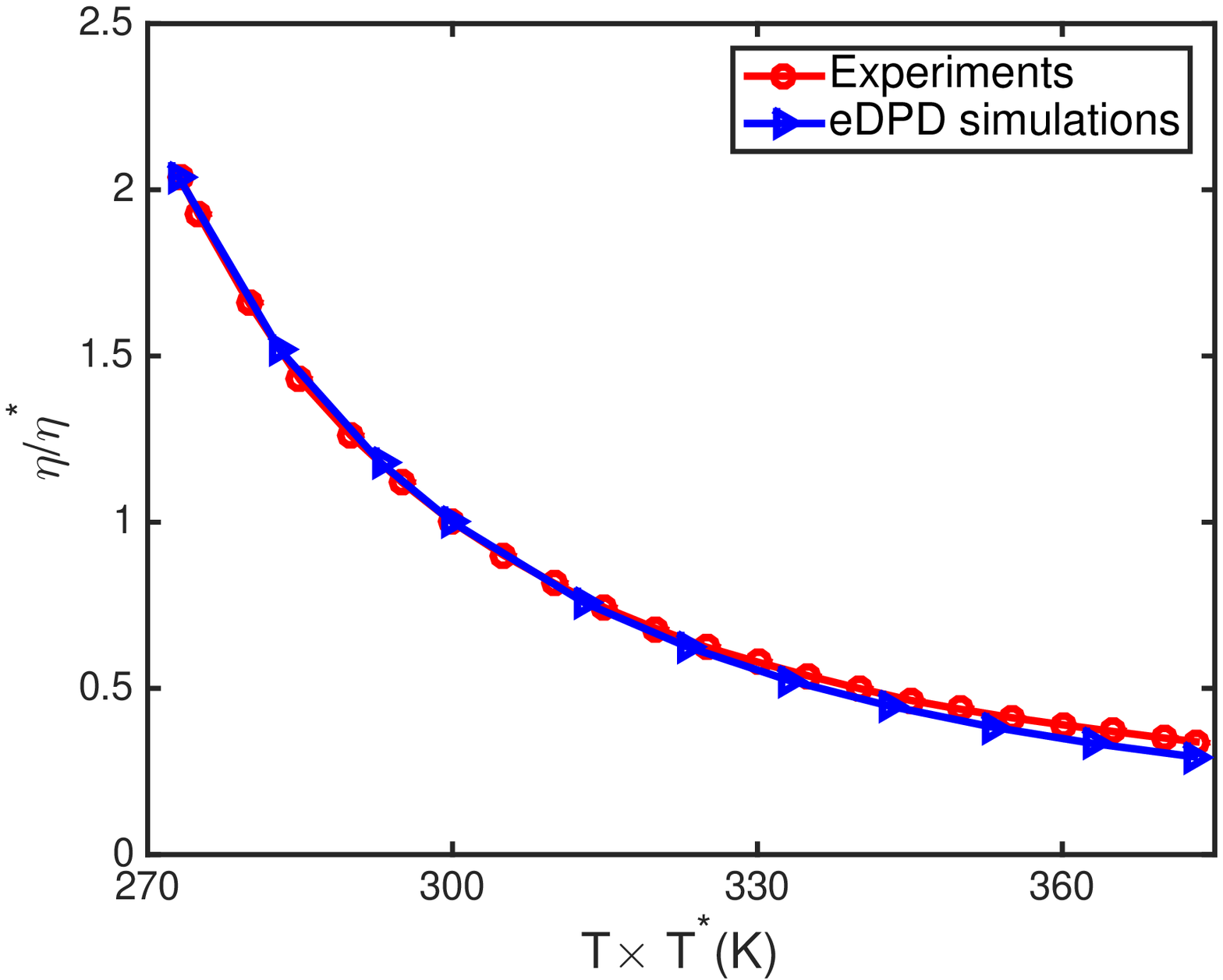}}
\caption{Comparison of temperature dependent (a) diffusivity and (b) viscosity
ranging from $273K$ to $373K$ between the experimental data of liquid water and
the results of eDPD simulations. Both the diffusivity and viscosity are scaled
by their value at $300K$.}
\label{fig:eDPD_comp}
\end{figure}

In this section we study the eDPD model for water with $s(T)$ given in Eq.
%-------------------------------------------------------------------------------
\eqref{eq:edpd_exp}. The model coefficients $C_i$ are set as
\begin{equation}
\begin{aligned}
 C_1(\xi_1) &= 3.8 + \sigma_{_{C_1}}\xi_1, & C_2(\xi_2) &= 0 + \sigma_{_{C_2}}\xi_2,\\
 C_3(\xi_3) &= 0 + \sigma_{_{C_3}}\xi_3, & C_4(\xi_4) &= 0 + \sigma_{_{C_4}}\xi_4,
\end{aligned}
\label{eq:para_edpd}
\end{equation}
%-------------------------------------------------------------------------------
where $(\sigma_{_{C_1}},\sigma_{_{C_2}},\sigma_{_{C_3}},\sigma_{_{C_4}})=
{(0.58,1.65,3.0,3.0)}$,
and $\bx = (\xi_1,\xi_2,\xi_3,\xi_4)$ are \textit{i.i.d} uniform random
variables distributed on $[-1,1]$. Therefore in this problem, we set
$\bm\theta=(C_1, C_2,C_3,C_4)$. We use a $4$th-order gPC ($70$ basis
functions) expansion to construct the surrogate model based on $50$ samples of
eDPD simulations. We aim to infer four parameters in the eDPD model given the
experiment data of diffusivity ($D$) and viscosity ($\eta$) of liquid water at
temperature $273K, 330K, 373K$, i.e., $T=(0.91,1.1,1.2433)$, hence to capture
the change of $D$ and $\eta$ as the temperature varies from $273K$ to $373K$.
Notice that $T,D,\eta$ are normalized by the corresponding value at temperature
$T^*=300K$ (see Section \ref{subsec:edpd_model}). Therefore, in this case,
$\bm P^t=(G_1,G_2,G_3,G_4,G_5,G_6)=(\eta_{0.91}, \eta_{1.1}, \eta_{1.2433}, D_{0.91}, D_{1.1}, D_{1.2433})$,
and the target property is
\begin{equation}  
\bm P^t=(2.040662, 0.579343, 0.339655, 0.4567 1.866233, 3.601906),
\label{eq:edpdtarget}
\end{equation}  
which are experimental data \cite{Bergman11,Holz00}. Figure \ref{fig:eDPD_infer} presents the
inference results for $C_i$ and each of them can be estimated by MAP method. In
this case, we do not observe the correlation between parameters by examining the
MCMC sampling points of pairwise parameters (not presented here). This is
because different parameter corresponds to different order of monomial
in the expression of $s(T)$, and each of them affect the quantity of interest
independently.
%-------------------------------------------------------------------------------
%-------------------------------------------------------------------------------
The set $\bm\theta$ we select is listed in Table \ref{tab:infer_eDPD}.
%-------------------------------------------------------------------------------
\begin{table}[!h]
\centering
\caption{Inferred parameters $\bm\theta$ for the liquid water.}
\begin{tabular}{C{4em}*{4}{C{6em}}}
\hline \hline
& $C_1$ & $C_2$ & $C_3$ & $C_4$ \\
\hline
$\bm\theta$ & $4.32$ & $-1.50$ & $1.41$ & $-1.68$\\
\hline \hline
\end{tabular}
\label{tab:infer_eDPD}
\end{table}
%-------------------------------------------------------------------------------
The comparisons of eDPD simulation {with} the experiment are presented in Figure
\ref{fig:eDPD_comp}. The eDPD simulation results match the experiment quite
well. Especially, the relative error of the viscosity is less than $2\%$ at each
temperature. The estimate of the diffusivity is also very accurate for
$273K\leq T\cdot T^* \leq 360K$, and the error grows slightly to around $5\%$ as
$T\cdot T^*$ approaches $373K$. These results demonstrate that with the present 
formulation we have improved the results in \cite{LiTLBK14} by introducing an 
advanced model for $s(T)$ and implementing the systematic approach to identify 
the parameters in the system.
%-------------------------------------------------------------------------------
%-------------------------------------------------------------------------------

%===============================================================================
%===============================================================================
%===============================================================================

\section{Summary and Discussion}
\label{sec:discussion}

In this work, we employ the gPC and compressive sensing method to
build accurate surrogate model efficiently for
mesoscopic polymer melt systems {and
non-isothermal liquid water}. We target various dynamic {or thermal}
properties and show that the numerical $L_2$ error of the gPC expansions
computed by the present method is comparable with (smaller than in some cases)
the sparse grid method. Moreover, different from the sparse grid method, the
present method enables us to incorporate new simulation data with no
restriction on the number of new samples, which is an advantage especially for
mesoscopic modeling and simulation, since in practice, the simulations are very
costly and may not be available at the specific sampling points in random
parameter space. On the other hand, compared with the Monte Carlo method, the
present method shows much faster convergence rate
(e.g., see \cite{DoostanO11, YangK13}) validating its
high efficiency in exploiting information from limited data. These results
demonstrate that the present method is well suited for studying uncertainty
quantification for mesoscopic models, in particular for those with {relatively} high
dimensional random parameter space, where only a limited number of simulations can
be afforded.

{Accurately recovering} the gPC coefficients of the target properties over the random
parameter space enables us to calibrate the model parameters with respect to the
observed target property {values (e.g., shear-rate viscosity, temperature-dependent 
viscosity and diffusivity, etc.)} \textcolor{black}{While a simple dumbbell FENE
model is considered in the current study for demonstration purposes, the present method
can be readily extended to more complex polymer melts and other soft matter models.
For example, as shown in Ref. \cite{Rafael_Soft_Matter_2016}, we may further 
incorporate the more general pressure tensor
by further introducing model variables such as tangential friction.} 
For models with high dimensional random
parameter space, the present work also provides a framework for model
optimization/reduction through identification of all possible parameter
degeneracies. We note that the dual effects from multiple parameters on target
properties of the mesoscopic models may not be easily identified on the basis of pure
theoretical modeling concepts. Systematic analysis on the intrinsic
relationship between those parameters relies on the full and accurate access to
the response surface of the targeted properties obtained from the present study.

Finally, we emphasize that several factors not considered in the present work
may further affect the performance of the proposed method. First, we note that the
present method relies on \textit{a prior} assumption that the
solution (gPC coefficients) is ``sparse'' in the gPC basis; otherwise, we may not be
able to obtain accurate results by means of compressive sensing. However, this
condition, in practice, is usually \textit{not that strong} for most mesoscopic
model systems. Given a system governed by a regular Hamiltonian formulation, the
physical properties can usually be well approximated by low orders of
polynomial function over parameter space. {Second, we emphasize that the
inferred parameter set for the mesoscopic model in the present
work is associated with the specific target properties we aim to recover. 
In the $6D$ polymer melt system in Sec. \ref{sec:6D_model}, we show that the
parameter degeneracy identified from the $6$ target properties also applies
to other dynamic properties. However, the general applicability of the
inferred parameter set and transferability to other properties remains an open
question. Essentially, the present framework enables to parameterize the
free energy space of mesoscopic model, on the other hand, extension to 
other properties further depends on the compatibility of the free energy space
with those properties.} \textcolor{black}{For example, to capture polydispersity and entanglement 
effects of long polymer chains, we need to start with some mesoscopic model 
where those structural properties can be (partially) characterized such that the 
present method can further facilitate finding the optimized model parameters.} 
Third, we consider
systems with small thermal noise in this paper. For systems with large thermal
noise, we may need to {adopt an appropriate method to separate model
parameter induced uncertainty from the intrinsic thermal fluctuation, e.g., see 
Ref. \cite{Maitre_Knio_2015}}
Finally, we note that the present work cannot be directly applied to study systems 
with parameter confidence range across the phase transition regime. Special treatment is needed
to take care of the abrupt discontinuity near the phase transition regime. \textcolor{black}{
Remarkably, the present method can still be applied if we are only interested in those 
properties which do not undergo drastic changes across the phase transition, e.g., morphological 
state under Rouse reptation transition for polymer melt system \cite{Reviewer_note}. 
} All
these issues need further study and will be addressed in future work.

%===============================================================================

\section{Appendix}
\textcolor{black}{
We provide a toy model further help readers to understand the procedure
in Algorithms \ref{alg:omp} and \ref{alg:omp_dpd}. Assume that we have a
polymer model $f(\xi_1,\xi_2)$ relying on two i.i.d. uniform random 
variables $\xi_1$ and $\xi_2$ with $\xi_1,\xi_2\sim\mathcal{U}[-1,1]$. Here $f$
takes the following form:
\begin{equation}
f(\xi_1,\xi_2) = \dfrac{20-3\sqrt{5}}{20} + \dfrac{7\sqrt{3}}{10}\xi_1 +
\dfrac{3\sqrt{5}}{8}\xi_1^2 + \dfrac{3}{50}\xi_1\xi_2 +
\dfrac{3\sqrt{5}}{40}\xi_2^2+\phi,
\end{equation}
where $\phi\sim\mathcal{N}(0,0.001)$ is a Gaussian random variable representing
the intrinsic noise. We now go through all the steps of Algorithms \ref{alg:omp} and
\ref{alg:omp_dpd} to demonstrate how they work. 
Since we will use Legendre polynomial expansion to approximate $f$, we first 
construct the two-dimensional normalized Legendre polynomial defined on 
$[-1,1]\times [-1,1]$ based on the one-dimensional normalized Legendre 
polynomials defined on $[-1,1]$ using tensor product rule. The one-dimensional
normalized Legendre polynomials up to second order are: 
$1, \sqrt{3}x, 2\sqrt{5}(3x^2-1)$, and the two-dimensional Legendre polynomials
are:
\begin{equation}
\begin{split}
&\psi_1(x_1,x_2)=1,~\psi_2(x_1,x_2)=\sqrt{3}x_1,~\psi_3(x_1,x_2)=\sqrt{3}x_2,\\
&\psi_4(x_1,x_2)=\dfrac{\sqrt{5}}{2}(3x_1^2-1),~\psi_5(x_1,x_2)=3x_1x_2,~
\psi_6(x_1,x_2)=\dfrac{\sqrt{5}}{2}(3x_2^2-1).
\end{split}
\end{equation}
Hence, the simple polymer model can be represented as:
\begin{equation}
\begin{split}
f(\xi_1,\xi_2) = & 1.0\psi_1(\xi_1,\xi_2) + 0.7\psi_2(\xi_1,\xi_2) +
0.05\psi_3(\xi_1,\xi_2) + 0.25\psi_4(\xi_1,\xi_2) \\ &
+ 0.01\psi_5(\xi_1,\xi_2) + 0.05\psi_6(\xi_1,\xi_2)+\phi.
\end{split}
\end{equation}
The coefficients vector is $\bm c=(1.0, 0.7, 0.05, 0.25, 0.01, 0.05)$. We will
show step by step how Algorithms \ref{alg:omp} and \ref{alg:omp_dpd} approximate
$\bm c$ based on the output of Monte Carlo simulations.
}

\textcolor{black}{
\noindent\textbf{Steps 1 and 2:} Run Monte Carlo simulation to obtain the output of the
polymer model. We generate five input samples of $\bm\xi=(\xi_1,\xi_2)$ and five
samples of $\phi$, then we can compute the five samples of $f$. We note that for
realistic problems, Monte Carlo simulation consists of only two parts: 
1) generating input samples of $\bm\xi$; 2) evaluating $f$ with samples of $\bm\xi$.
Table \ref{tab:append_mc} lists input samples $\bm\xi^1,\bm\xi^2,\cdots,\bm\xi^5$ and
corresponding output samples $f^1,,f^2,\cdots,f^5$.
\begin{table}[h]
\caption{Samples of $\bm\xi$ and corresponding samples of $f$}
\begin{tabular}{C{3em}*{5}{|C{8em}}}
\hline\hline
$\bm\xi$ & (0.0258,-0.0790) & (-0.2992,-0.8099) & (-0.1327,0.4185) & (-0.7681,-0.8438) & (-0.2615,-0.9327) \\
\hline
$f$ & 0.6905 &  0.4235 &  0.5822 &  0.2942 &  0.4780\\
\hline\hline
\end{tabular}
\label{tab:append_mc}
\end{table}
}

\textcolor{black}{
\noindent\textbf{Step 3:} We construct the ``measurement matrix" $\newtensor\Psi$
by evaluating two-dimensional Legendre polynomial $\psi_i, i=1,\cdots,6$ at the
samples of $\bm\xi$. More precisely, $\Psi_{ij}=\psi_j(\bm\xi^i)$, where
$i=1,2,\cdots,5$ and $j=1,2,\cdots,6$. Hence, we obtain the following linear
system:
\begin{equation}
\begin{pmatrix}
1.0000 &   0.0447 &  -0.1369 &  -1.1158 &  -0.0061 &  -1.0971 \\
1.0000 &  -0.5182 &  -1.4028 &  -0.8178 &   0.7270 &   1.0821 \\ 
1.0000 &  -0.2298 &   0.7248 &  -1.0590 &  -0.1665 &  -0.5307 \\ 
1.0000 &  -1.3303 &  -1.4616 &   0.8606 &   1.9443 &   1.2703 \\ 
1.0000 &  -0.4529 &  -1.6156 &  -0.8887 &   0.7317 &   1.8001 \\
\end{pmatrix}
\begin{pmatrix}
c_1\\ c_2\\ c_3\\ c_4\\ c_5
\end{pmatrix}=
\begin{pmatrix}
 0.6905 \\ 0.4235 \\  0.5822 \\  0.2942 \\ 0.4780
\end{pmatrix} + \vec\varepsilon.
\end{equation}
}

\textcolor{black}{
\noindent\textbf{Step 4:} In the demonstration we choose a threshold
$\delta=0.05$. In practice, it is estimated by cross-validation and details can
be found in \cite{DoostanO11}.
}

\textcolor{black}{
\noindent\textbf{Step 5:} Solve the $\ell_0$ minimization by using Algorithm 1.
The step by step details are as follows:
\begin{enumerate}
\item $k=0,\bm c^0=(0,0,0,0,0,0),\bm r^0=(0.6905,0.4235,0.5822,0.2942,0.4780)^T,
  \mathcal{S}^0=\emptyset$. Then $\epsilon$ in Algorithms \ref{alg:omp} are: 
 $(\epsilon(1),\epsilon(2),\epsilon(3),\epsilon(4),\epsilon(5),\epsilon(6))=(0.0916,0.9337,1.0130,0.5154,1.0488,1.2580)$.
 Hence, we select $j_0=1$, and $\mathcal{S}^1=\mathcal{S}^0 \cup \{1\}=\{1\},\bm
 r^1=(0.1969,-0.0702,0.0886,-0.1995.-0.0157)^T$. 
 Here $\Vert\bm r^1\Vert_2=0.3026$ which is larger than the threshold $\delta$, so we need to
 continue expanding $\mathcal{S}$.
\item $k=1,\bm c^1=(0.4937,0,0,0,0,0),\bm r^1=(0.1969,-0.0702,0.0886,-0.1995.-0.0157)^T,
  \mathcal{S}^1=\{1\}$. Then $\epsilon$ in Algorithms \ref{alg:omp} are:
$(\epsilon(2),\epsilon(3),\epsilon(4),\epsilon(5),\epsilon(6))=(0.0531,0.0633,0.0541,0.0469,0.0403)$.
Hence, we select $j_0=6$, and $\mathcal{S}^2=\mathcal{S}^1 \cup \{6\}=\{1,6\}$.
Then, we compute the minimizer of $\Vert\newtensor\Psi\bm c-\bm b\Vert_2^2$ subject to
$\text{Support}\{\bm c\}=\mathcal{S}^2=\{1,6\}$. The results is 
$\bm c^2=(0.5439,0,0,0,0,-0.0995)$ and the residual $\bm
r^2=(-0.0374,0.0128,0.0145,0.1233,-0.1132)$. Since $\Vert\bm r^2\Vert_2$ is
still larger than the threshold $\delta$, we continue expanding $\mathcal{S}$.
\item After five similar steps, we obtain $\mathcal{S}^5=\{1,2,3,4,6\}, \bm
c^5=(1.0120,0.7045,0.0477,0.2598,0,0.0516)$ and $\Vert r^5\Vert_2<\delta$, hence
we can stop.
\end{enumerate}
Finally, we use $\bm c^5$ to approximate the coefficients $\bm c$ in the polymer
model $f$.
}

%===============================================================================
%===============================================================================

\section{Acknowledgment}
This research is sponsored by the Army Research Laboratory and was accomplished
under Cooperative Agreement Number W911NF-12-2-0023 to University of Utah. We
also acknowledge partial support from the new Collaboratory on Mathematics for
Mesoscopic Modeling of Materials (CM4) supported by DOE. We would like to thank
Hui Wang, Zhongqiang Zhang, Mingge Deng, Xiaoxing Cheng \textcolor{black}{and two
anonymous reviewers}
for helpful discussions and \textcolor{black}{constructive suggestions}.

\bibliographystyle{elsarticle-num}
\bibliography{reference}
%%%%%%%%%%%%%%%%%%%%%%%%%%%%%%%%%%%%%%%%%%%%%%%%%%%%%%%%%%%%%%%%%%%%%%%%%%%%%%%%%%%%%%%%%%%%%%%%%%
\end{document}